\newcommand{\PRE}[1]{{#1}} 
\documentclass[12pt,aps,prd,preprint,tightenlines,superscriptaddress,
amsmath,amssymb,floatfix,
nofootinbib]{revtex4}

\usepackage{slashed}
\usepackage{enumerate}
\usepackage{color}
\usepackage[pdftex]{graphicx}
\usepackage[colorlinks=true,citecolor=blue,urlcolor=blue]{hyperref}
\usepackage[utf8]{inputenc}

\graphicspath{{paperfigures/}}	        

\newcommand{\OmegaDM}{\Omega_{\text{DM}}}

\newcommand{\ifb}{\,\text{fb}^{-1}}

\newcommand{\gev}{\text{GeV}}
\newcommand{\tev}{\text{TeV}}

\newcommand{\cm}{\text{cm}}

\newcommand{\s}{\text{s}}

\newcommand{\cSI}{\text{SI}}
\newcommand{\cSD}{\text{SD}}

\newcommand{\eg}{{\em e.g.}}

\newcommand{\Eqref}[1]{Equation~(\ref{#1})}
\renewcommand{\eqref}[1]{Eq.~(\ref{#1})}
\newcommand{\eqsref}[2]{Eqs.~(\ref{#1}) and (\ref{#2})}
\newcommand{\secref}[1]{Sec.~\ref{sec:#1}}
\newcommand{\secsref}[2]{Secs.~\ref{sec:#1} and \ref{sec:#2}}
\newcommand{\figref}[1]{Fig.~\ref{fig:#1}}

\newcommand{\tableref}[1]{Table~\ref{table:#1}}

\newcommand{\software}[1]{\textsc{#1}}
\newcommand{\micromegas}{\textsc{micrOMEGAs}}

\newcommand{\abs}[1]{\ensuremath{\left|#1\right|}}

\begin{document}

\preprint{UCI--TR--2016--08}

\title{\PRE{\vspace*{01.0in}}
Heavy Bino Dark Matter and Collider Signals \\
in the MSSM with Vector-like 4th-Generation Particles 
\PRE{\vspace*{.5in}}}

\author{Mohammad Abdullah\footnote{maabdull@uci.edu}}
\affiliation{Department of Physics and Astronomy, University of
  California, Irvine, California 92697, USA
\PRE{\vspace*{.2in}}}

\author{Jonathan L.~Feng\footnote{jlf@uci.edu}}
\affiliation{Department of Physics and Astronomy, University of
  California, Irvine, California 92697, USA
\PRE{\vspace*{.2in}}}

\author{Sho Iwamoto\footnote{sho@physics.technion.ac.il}}
\affiliation{Physics Department, Technion---Israel Institute of 
Technology, Haifa 32000, Israel
\PRE{\vspace*{.4in}}}

\author{Benjamin Lillard\footnote{blillard@uci.edu} \PRE{\vspace*{.2in}}}
\affiliation{Department of Physics and Astronomy, University of
  California, Irvine, California 92697, USA
\PRE{\vspace*{.2in}}}


\begin{abstract}
\PRE{\vspace*{.2in}} MSSM4G models, in which the minimal supersymmetric standard model is extended to include vector-like copies of standard model particles, are promising possibilities for weak-scale supersymmetry.  In particular, two models, called QUE and QDEE, realize the major virtues of supersymmetry (naturalness consistent with the 125 GeV Higgs boson, gauge coupling unification, and thermal relic neutralino dark matter) without the need for fine-tuned relations between particle masses.  We determine the implications of these models for dark matter and collider searches.  The QUE and QDEE models revive the possibility of heavy Bino dark matter with mass in the range 300--700 GeV, which is not usually considered.  Dark matter direct detection cross sections are typically below current limits, but are naturally expected above the neutrino floor and may be seen at next-generation experiments. Indirect detection prospects are bright at the Cherenkov Telescope Array, provided the 4th-generation leptons have mass above 350 GeV or decay to taus.  In a completely complementary way, discovery prospects at the LHC are dim if the 4th-generation leptons are heavy or decay to taus, but are bright for 4th-generation leptons with masses below 350 GeV that decay either to electrons or to muons.  We conclude that the combined set of direct detection, CTA, and LHC experiments will discover or exclude these MSSM4G models in the coming few years, assuming the Milky Way has an Einasto dark matter profile.
\end{abstract}

\pacs{Fourth generation particles, supersymmetric models, dark matter, collider physics}

\maketitle

\section{Introduction}
\label{sec:intro}

Weak-scale supersymmetry (SUSY) has the potential to solve the gauge hierarchy problem, accommodate grand unification, and explain dark matter in the form of a thermal relic neutralino.  This potential has been sullied a bit by the lack of superpartners at the LHC and the measured Higgs mass of 125 GeV, which is higher than typically expected in the minimal supersymmetric standard model (MSSM) with sub-TeV superpartners.  There remain, however, particular versions and extensions of the MSSM that preserve some or all of SUSY's potential virtues, while remaining viable, even in light of LHC data~\cite{Feng:2013pwa,Craig:2013cxa}.  

MSSM4G models are extensions of this kind.  In MSSM4G models, the MSSM is extended to include 4th- (and, possibly, 5th-) generation vector-like copies of standard model (SM) particles.  The new generations contribute to the Higgs boson mass, raising it to 125 GeV without needing to raise any particle masses above $2~\tev$, preserving naturalness~\cite{Moroi:1991mg,Moroi:1992zk}. At the same time, if the new multiplets are judiciously chosen, gauge coupling unification is preserved.  In fact, requiring perturbative gauge coupling unification reduces the number of MSSM4G models to two, the QUE and QDEE models~\cite{Martin:2009bg}.  Last, the new fermions provide new annihilation channels for thermal relic neutralinos, opening up qualitatively new possibilities for Bino dark matter.  As discussed in Ref.~\cite{Abdullah:2015zta}, the annihilation process  to 4th- (and 5th-) generation isosinglet charged leptons, $\tilde{B} \tilde{B} \to \tau_{4,5}^+ \tau_{4,5}^-$, is remarkably efficient, because it is enhanced by the large hypercharge factor $(Y_{\tau_{4,5}})^4 = 16$ and is not chirality-suppressed by small fermion masses.  As a result, this process may single-handedly dominate the combined effect of tens of MSSM annihilation channels, reviving the viability of 300--700 GeV Bino dark matter, which, barring coannihilation, overcloses the Universe in the MSSM. 

In this study, we determine the prospects for discovering MSSM4G models at dark matter and collider experiments.  The possibility of heavy Binos with the correct thermal relic density is not realized in the MSSM, and so is not very well studied.  As we will see, for direct detection, the scattering of Binos is highly suppressed, first by Yukawa couplings, as is typical of ``Higgs-mediated" dark matter candidates, but, second, also by the smallness of the Higgsino component of the dark matter. The predicted cross sections are typically below current bounds, but are above the neutrino floor, making MSSM4G dark matter ideal targets for future searches.  In short, direct detection eliminated ``$Z$ models" long ago, are currently exploring ``Higgs models," and will soon probe these ``Bino models" on their way to the neutrino floor.

For indirect detection, MSSM4G dark matter annihilates to 4th-generation leptons, which then decay to $W$, $Z$, and $h$ bosons and SM leptons.  We examine the prospects for detecting these decays through charged particles, neutrinos, and gamma rays, and find particularly promising prospects for future gamma-ray experiments, such as the Cherenkov Telescope Array (CTA), when the 4th-generation leptons are heavy or when they decay to taus.

Last, we examine the prospects for colliders.  In contrast to the conclusions for CTA, the LHC is most promising when the 4th-generation leptons are light and decay either to electrons only or muons only.  When they are heavy or decay to taus, even the high luminosity LHC with $3~\text{ab}^{-1}$ cannot discover new particles in the parameter region favored by thermal relic density constraints~\cite{Kumar:2015tna}.  The LHC and CTA regions of sensitivity are therefore highly complementary. We also consider the prospects for TeV-scale lepton colliders, such as the International Linear Collider (ILC).

This study is organized as follows.  In \secref{models} we present the QUE and QDEE models in detail.  We then consider discovery signals in direct detection, indirect detection, and colliders in Secs.~\ref{sec:direct}, \ref{sec:indirect}, and \ref{sec:colliders}, respectively, and present our conclusions in \secref{conclusions}.

\section{The QUE and QDEE Models}
\label{sec:models}

The MSSM contains three generations of $\hat{Q}$, $\hat{U}$, $\hat{D}$, $\hat{L}$, and $\hat{E}$ matter superfields, which are the quark isodoublet, up-type quark isosinglet, down-type quark isosinglet, lepton isodoublet, and charged lepton isosinglet superfields, respectively.  In MSSM4G models, these are supplemented by vector-like copies of SM fermions, that is, both left- and right-handed versions of fermions whose SU(3)$\times$SU(2)$\times$U(1)$_Y$ charges are identical to those of SM fermions, along with their scalar superpartners.  Vector-like matter preserves anomaly cancellation and typically satisfies electroweak precision constraints when the vector-like mass contributions dominates those from Yukawa terms.

As noted above, requiring the extra field content to both preserve gauge coupling unification and naturally raise the Higgs boson mass to 125 GeV restricts the possible MSSM4G models to two: the QUE and QDEE models.  The line of reasoning leading to this remarkable conclusion has been detailed elsewhere; see, \eg, Ref.~\cite{Martin:2009bg}.  Rather than repeating the argument here, in this section we simply define the models and specify the simplifying assumptions we adopt to reduce the number of parameters to a manageable number. Our approach, including the notation and conventions, follows Ref.~\cite{Abdullah:2015zta}.

\subsection{The QUE Model}

In the QUE model, we add vector-like 4th-generation copies of the $\hat{Q}$, $\hat{U}$, and $\hat{E}$ superfields, equivalent to adding superfields in the ${\bf 10} + {\bf \overline{10}}$ representations of SU(5).  The additional particles in the QUE model are
\begin{eqnarray}
\text{Dirac (4-component) fermions:} && T_4, B_4, t_4, \tau_4 \\
\text{Complex scalars:} && \tilde{T}_{4L}, \tilde{T}_{4R}, 
\tilde{B}_{4L}, \tilde{B}_{4R}, \tilde{t}_{4L}, \tilde{t}_{4R}, 
\tilde{\tau}_{4L}, \tilde{\tau}_{4R} \ ,
\end{eqnarray}
where the subscripts 4 denote 4th-generation particles, upper- and
lower-case letters denote isodoublets and isosinglets, respectively,
and $L$ and $R$ denote scalar partners of left- and right-handed
fermions, respectively.  This notation exploits the fact that none of the models we consider contains isodoublet leptons, avoiding the need for upper-case taus.

The SUSY-preserving interactions are
specified by the superpotential
 \begin{equation}
W_{\text{QUE}} = 
M_{Q_4}\hat{Q}_4 \hat{\bar{Q}}_4 + M_{t_4} \hat{t}_4 \hat{\bar{t}}_4 
+ M_{\tau_4} \hat{\tau}_4 \hat{\bar{\tau}}_4 
+ k \hat{H}_u \hat{Q}_4 \hat{\bar{t}}_4 
- h \hat{H}_d \hat{\bar{Q}}_4 \hat{t}_4 \ ,
\label{sprptntl}
\end{equation}
where $\hat{Q}_4 = (\hat{T}_4,
\hat{B}_4)$ is the quark isodoublet, $\hat{t}_4$ and $\hat{\tau}_4$
are the quark and lepton isosinglets, and the vector-like masses
$M_{Q_4}$, $M_{t_4}$, and $M_{\tau_4}$ and the Yukawa couplings $k$
and $h$ are all free parameters. The 4th-generation fermions must also mix with MSSM fields so that they can decay and are not cosmologically troublesome.  We will assume small, but
non-vanishing, mixings of these fermions that are dominantly to either the 1st-, 2nd-, or 3rd-generation
fermions.  Which generation it is has little impact on the direct detection signals we discuss, but is highly relevant for the indirect detection and collider signals, as we will see in \secsref{indirect}{colliders}.  Finally, there are the soft SUSY-breaking terms
\begin{eqnarray}
\mathcal{L}_{\text{QUE}} &=& 
- m^2_{\tilde{Q}_4} \lvert \tilde{Q}_4 \rvert^2
- m^2_{\tilde{\bar{Q}}_4} \lvert \tilde{\bar{Q}}_4 \rvert^2
- m^2_{\tilde{t}_4} \lvert \tilde{t}_4 \rvert^2 
- m^2_{\tilde{\bar{t}}_4}\lvert \tilde{\bar{t}}_4 \rvert^2 
- m^2_{\tilde{\tau}_4}\lvert \tilde{\tau}_4 \rvert^2 
- m^2_{\tilde{\bar{\tau}}_4}\lvert \tilde{\bar{\tau}}_4 \rvert^2 
\nonumber \\ 
&& - kA_{t_4} H_u \tilde{Q}_4 \tilde{\bar{t}}_4 
+ hA_{b_4} H_d \tilde{\bar{Q}}_4 \tilde{t}_4
- B_{Q_4}\tilde{Q}_4\tilde{\bar{Q}}_4 
- B_{t_4}\tilde{t}_4 \tilde{\bar{t}}_4 
- B_{\tau_4}\tilde{\tau}_4 \tilde{\bar{\tau}}_4 \ ,
\end{eqnarray} 
where all the coefficients are free, independent parameters.

To reduce the number of parameters to a reasonable number, we make some simplifying assumptions about the weak-scale values of
these parameters.  To maximize the
radiative corrections to the Higgs mass from the 4th-generation quark sector, we fix the
up-type Yukawa coupling to be at its quasi-fixed point value
$k=1.05$~\cite{Martin:2009bg}.  The down-type Yukawa coupling can also increase the Higgs mass slightly for $h < 0$, but the effect is small for moderate and large $\tan\beta$, and so we set $h = 0$ for simplicity.  We choose the 4th-generation $A$-parameters such that
there is no left--right squark mixing, that is, $A_{t_4} - \mu \cot
\beta = 0$, $A_{b_4} - \mu \tan \beta = 0$, and that the 4th-generation $B$-parameters are negligible, ignoring the $CP$-phases as well.

For the masses, we assume spectra of the extra fermions and
sfermions that can be specified by 4 parameters: the unified
(weak-scale) squark, slepton, quark, and lepton masses
\begin{equation}
 \begin{split}
  m_{\tilde{q}_4} &\equiv m_{\tilde{T}_{4L}} = m_{\tilde{T}_{4R}} 
= m_{\tilde{B}_{4L}} = m_{\tilde{B}_{4R}} = m_{\tilde{t}_{4L}} 
= m_{\tilde{t}_{4R}} \\
m_{\tilde{\ell}_4} &\equiv m_{\tilde{\tau}_{4L}} 
= m_{\tilde{\tau}_{4R}} \\
m_{q_4} &\equiv m_{T_4} = m_{B_4} = m_{t_4} \\
m_{\ell_4} &\equiv m_{\tau_4} \ .
 \end{split}
 \label{eq:unifiedmass-QUE}
\end{equation}

Finally, we assume $|\mu|$ is greater than the Bino mass $M_1$, so that the lightest neutralino is Bino-like with a small Higgsino admixture, and, in a slight abuse of notation, we denote this lightest neutralino as $\tilde{B}$.  We further assume that the Bino is the lightest supersymmetric
particle (LSP), but heavier than some 4th-generation fermions, so that it
can annihilate to them and reduce its thermal relic density.  For
simplicity, we assume the mass ordering
\begin{equation}
m_{\tilde{q}_4}, m_{\tilde{\ell}_4}, m_{q_4}, |\mu|
> m_{\tilde{B}} > m_{\ell_4}  \ ,
\label{QUEparameters}
\end{equation}
so that Binos annihilate to 4th-generation leptons, but not to 4th-generation quarks.  As  noted above, the addition of the 4th-generation
lepton channels is enough to reduce the Bino relic density to allowed
levels.  This ordering also allows the new colored particles to be
heavy enough to avoid LHC bounds and contribute sufficiently to the Higgs mass correction.

The relic density constraints were investigated in Ref.~\cite{Abdullah:2015zta} and we summarize the results here. As noted above, because the $\tilde{B} \tilde{B} \to \tau_4^+ \tau_4^-$ process is enhanced by large hypercharges and not chirality-suppressed by small fermion masses, it dominates the annihilation cross section.  The $S$-wave and $P$-wave pieces of the thermally-averaged annihilation cross section were derived and used to calculate the relic density as a function of $m_{\tilde B}$, $m_{\ell_4}$, and $m_{\tilde{\ell}_4}$. We did not include coannihilation, and so the validity of our calculation was restricted to regions where Binos and sleptons were not degenerate to more than 5\%. In addition, we avoided regions with degenerate Binos and leptons, where the partial wave expansion breaks down. 

It was shown that $m_{\tilde B}$ can be increased up to about $540~\gev$ without overclosing the Universe, as long as $m_{\ell_{4}}$ is not much lower and $m_{\tilde{\ell}_{4}}$ is not much higher than $m_{\tilde B}$. The required lepton and slepton masses have not been excluded by experiments.  The lower bounds on the Bino mass from collider searches are less definitive since they make assumptions about other superpartners. We will consider the masses $m_{\tilde B} > 200~\gev$ and be particularly interested in masses above 300 GeV, which is not very well-studied, since Binos, in the absence of coannihilation effects, overclose the Universe for these masses in the MSSM. 

In summary, the relevant parameters of the QUE model are those listed in \eqref{QUEparameters}, along with $\tan\beta$, the CP-odd Higgs mass $m_A$, and the masses of the MSSM superpartners.  The prospects for direct detection also depend somewhat on the masses of the MSSM squarks, as we will discuss in \secref{direct}. We take their left--right mixing to be small, and we consider the mass ranges shown below in \tableref{parambounds}. As the direct detection cross section is dominated by Higgsino scattering, the effects of $\mu$ and $\tan\beta$ are more important than the squark masses.

\subsection{The QDEE Model}

If one drops the GUT multiplet requirement, there is another
possibility consistent with gauge coupling
unification and a natural 125 GeV Higgs mass~\cite{Martin:2009bg}: the QDEE model, with the $U$ of the QUE model replaced by a $D$, and an additional (5th-generation) $E$.
With notation similar to that above, the QDEE model has the extra
particles
\begin{eqnarray}
\text{Dirac (4-component) fermions:} && T_4, B_4, b_4, \tau_4 , \tau_5 \\
\text{Complex scalars:} && \tilde{T}_{4L}, \tilde{T}_{4R}, 
\tilde{B}_{4L}, \tilde{B}_{4R}, \tilde{b}_{4L}, \tilde{b}_{4R}, 
\tilde{\tau}_{4L}, \tilde{\tau}_{4R} , 
\tilde{\tau}_{5L}, \tilde{\tau}_{5R} \ .
\end{eqnarray}

The superpotential is
\begin{equation}
W_{\text{QDEE}} = M_{Q_4} \hat{Q}_4 \hat{\bar{Q}}_4 
+ M_{b_4} \hat{b}_4 \hat{\bar{b}}_4 
+ M_{\tau_4} \hat{\tau}_4 \hat{\bar{\tau}}_4
+ M_{\tau_5} \hat{\tau}_5 \hat{\bar{\tau}}_5 
+ k \hat{H}_u \hat{Q}_4 \hat{\bar{b}}_4 
- h \hat{H}_d \hat{\bar{Q}}_4 \hat{b}_4 \ ,
\label{sprptntl2}
\end{equation}
and the soft SUSY-breaking terms are
\begin{eqnarray}
\mathcal{L}_{\text{QDEE}} \! \! &=& 
\! \! - m^2_{\tilde{Q}_4} \lvert \tilde{Q}_4 \rvert^2
\! \! - \! m^2_{\tilde{\bar{Q}}_4} \lvert \tilde{\bar{Q}}_4 \rvert^2
\! \! - \! m^2_{\tilde{b}_4} \lvert \tilde{b}_4 \rvert^2 
\! \! - \! m^2_{\tilde{\bar{b}}_4} \lvert \tilde{\bar{b}}_4 \rvert^2 
\! \! - \! m^2_{\tilde{\tau}_4} \lvert \tilde{\tau}_4 \rvert^2 
\! \! - \! m^2_{\tilde{\bar{\tau}}_4} \lvert \tilde{\bar{\tau}}_4 \rvert^2
\! \! - \! m^2_{\tilde{\tau}_5} \lvert \tilde{\tau}_5 \rvert^2
\! \! - \! m^2_{\tilde{\bar{\tau}}_5} 
   \lvert \tilde{\bar{\tau}}_5 \rvert^2 \nonumber \\
&& \! \! - k A_{t_4} H_u \tilde{Q}_4 \tilde{\bar{b}}_4 
+ h A_{b_4} H_d \tilde{\bar{Q}}_4 \tilde{b}_4 
 - B_{Q_4} \tilde{Q}_4 \tilde{\bar{Q}}_4 
 - B_{b_4} \tilde{b}_4 \tilde{\bar{b}}_4 
 - B_{\tau_4} \tilde{\tau}_4 \tilde{\bar{\tau}}_4 
 - B_{\tau_5} \tilde{\tau_5} \tilde{\bar{\tau}}_5 \ .
\end{eqnarray} 

For simplifying assumptions, as in the QUE models, we set the up-type Yukawa coupling to its quasi-fixed point value $k = 1.047$, the down-type Yukawa coupling to $h=0$, the 4th- and 5th-generation $A$-parameters to eliminate left--right squark mixing, and we assume negligible 4th- and 5th-generation $B$-parameters.  For the QDEE model masses, we also assume 4 unifying masses
\begin{equation}
 \begin{split}
 m_{\tilde{q}_4} &\equiv m_{\tilde{T}_{4L}} = m_{\tilde{T}_{4R}} 
 = m_{\tilde{B}_{4L}} = m_{\tilde{B}_{4R}} 
 = m_{\tilde{b}_{4L}} = m_{\tilde{b}_{4R}} \\
 m_{\tilde{\ell}_4} &\equiv m_{\tilde{\tau}_{4L}} = m_{\tilde{\tau}_{4R}}
 = m_{\tilde{\tau}_{5L}} = m_{\tilde{\tau}_{5R}} \\
 m_{q_4} &\equiv m_{T_4} = m_{B_4} = m_{b_4} \\
 m_{\ell_4} &\equiv m_{\tau_4} = m_{\tau_5} \ ,
\end{split}
\label{eq:unifiedmass-QDEE}
\end{equation}
and the same ordering of masses given in \eqref{QUEparameters}.  As in the QUE model, the relevant parameters are these masses, $|\mu|$, $m_A$, $\tan\beta$, and the masses of the MSSM superpartners.

Given the extra lepton generation, the Bino annihilation cross section is twice as efficient in the QDEE model as in the QUE model, allowing for Bino masses of up to 740 GeV. This can be understood by observing that $\langle \sigma v \rangle \sim m^{-2}$, which implies that the allowed Bino mass should be larger than 540 GeV by a factor of about $\sqrt{2}$. 

The mass ranges of the Bino, 4th generation fields and the MSSM stop in both models are summarized in Table~\ref{table:paramrange}. Note that the relic density and Higgs mass requirements  lead to correlations between the values of some of these parameters. See Ref.~\cite{Abdullah:2015zta} for details.

\begin{table}[tb]
\begin{tabular}{| c | c | c |}\hline
Parameter	 & QUE (\gev)	& QDEE (\gev) \\ \hline
$M_{\tilde{B}}$ & $200-540$ & $200-740$ \\ \hline
$m_{\tilde{q}_4}$ & $1000-4000$ &$ 1000-4000$ \\ \hline
$m_{\tilde{\ell}_4}$ & $350-550$ & $400 - 750$ \\ \hline 
$m_{q_4}$ & $1000-2000$ & $1000-2000$ \\ \hline
$m_{\ell_4}$ & $170 - 450$ & $170 - 620$ \\ \hline
$m_{\tilde{t}}$ & $1000-4000$ & $1000-4000$ \\ \hline
\end{tabular}
\caption{The mass ranges we consider for the parameters in \eqsref{eq:unifiedmass-QUE}{eq:unifiedmass-QDEE}} as well as the MSSM stop. 
\label{table:paramrange}
\end{table}

\section{Direct Detection of Dark Matter}
\label{sec:direct}

In both the QUE and QDEE models, the lightest neutralino $\tilde{B}$ may interact strongly enough with nuclear matter to be detected by current or future direct detection experiments. We explore this possibility for both spin-independent (\cSI) and spin-dependent (\cSD) direct detection. In \secref{dd:a} we discuss the qualitative behavior we expect, given an approximate analytic expression for the effective couplings between neutralinos and nucleons, which is derived in Appendix~\ref{sec:ddderivation}. In \secsref{SI}{SD} we use the \micromegas\ package to calculate the \cSI\ and \cSD\ cross sections for a wide range of parameter space and compare these predictions against current and future experimental sensitivities. 

\subsection{Effective Neutralino--Nucleon Coupling}
\label{sec:dd:a}

Interactions between Bino dark matter and the nucleons of a particle detector are primarily mediated by $t$-channel scalar Higgses, $ h^0$ and $ H^0$, or by $s$-channel squarks, $\tilde q_i$. As a result, the QUE and QDEE have nearly identical direct detection prospects: the contribution from the 4th-generation quarks is limited to $\tilde{B} \tilde{B} gg$ interactions mediated by heavy squark/quark loops, which we neglect.

The non-observation of squarks at the LHC suggests that their masses are significantly larger than the Higgs mass. SI squark-mediated scattering is proportional to left--right squark mixing angles, which are highly suppressed by quark masses for the most relevant quarks, namely those of the first and second generation. For $\mathcal O(\tev)$ squark masses we find that the \cSI\ cross section is dominated by Higgs-mediated scattering, despite the associated suppression by Yukawa couplings and the small Higgsino fraction of the neutralino.  In Appendix~\ref{sec:ddderivation} we derive a simple expression for the effective neutralino--nucleon coupling for Bino-like neutralinos, in the limit of large squark masses and moderate-to-large values of $\tan\beta$ ($5 \leq \tan\beta \leq 50$). The main results are given in this section.

The differential cross section for dark matter scattering from a nucleus with mass number $A$ and charge $Z$ is~\cite{Drees:1993bu}
\begin{equation}
\frac{d\sigma}{d\abs{\vec{q}}^2} = \frac{1}{\pi v^2} \left[ Z f_p + (A-Z) f_n \right]^2 F^2(Q) \ ,
\end{equation}
where $\vec{q}$ is the momentum transferred in the interaction; $v$ is the velocity of the dark matter; $f_p$ and $f_n$ are the effective couplings to protons and neutrons, respectively; and $F(Q)$ is the nuclear form factor, where $Q$ is the energy transfer. In our model, $f_p$ and $f_n$ tend to be approximately equal.

When all the squarks and the heavy neutral Higgs boson are significantly heavier than the light Higgs boson with mass $m_h = 125~\gev$ and $\tan\beta$ is moderate or large, the couplings of the dark matter are approximately 
\begin{eqnarray}
\frac{f_{p,n}}{m_{p}} &\approx & N_{41} \left[ N_{21} - N_{11} \tan\theta_W \right] \frac{ g^2 }{4m_W m_h^2 }  \left[f_{Td} -f_{Tu} +  f_{Ts} - \frac{2}{27} f_{TG} \right] \ ,
\label{eq:fpeff}
\end{eqnarray} 
where the coefficients $N_{j1}$ are the components of the neutralino dark matter in the gauge basis $\{\tilde{B}, \tilde{W}, \tilde{H}_d, \tilde{H}_u\}$, $\theta_W$ is the weak mixing angle, and $f_{Tq}$ and $f_{TG}$ parameterize the quark and gluon content of the nucleon.  For Bino-like dark matter, $N_{11} \sim 1$, and the other coefficients are suppressed by powers of $M_1 / \mu$.  Expanding for large $|\mu|$, we find
\begin{eqnarray}
\frac{f_{p,n}}{m_p} &=& \frac{M_1 m_Z \tan\theta_W \sin\theta_W}{\mu^2 - M_1^2 + m_Z^2 \sin^2\theta_W } \left( \frac{g^2}{4 m_W m_h^2} \right) \left[ f_{Td} -f_{Tu} +  f_{Ts} - \frac{2}{27} f_{TG} \right]. \label{eq:finalcoupling} 
\end{eqnarray}

Values for $f_{Tu}$ and $f_{Td}$ can be obtained from pion-nucleon scattering; $f_{Ts}$ is found more precisely from lattice calculations. The sum $f_{TG} + \sum_{u,d,s} f_{Tq} = 1$ determines $f_{TG}$.  In \micromegas, the following values are used for $f_{Tq}$~\cite{Belanger:2014vza}:
\begin{align}
f_{Tu}^{(p)} = 0.0153 &\, , & f_{Td}^{(p)} = 0.0191 &\, ,& f_{Tu}^{(n)} = 0.011 &\, ,& f_{Td}^{(n)} = 0.0273 &\, ,& f_{Ts}^{(n,p)} = 0.0447\, .
\label{eq:scalarquark}
\end{align} 
The value for $f_{Ts}$ agrees with recent lattice calculations~\cite{Junnarkar:2013ac}, which find $f_{Ts} = 0.053 \pm 0.011 \pm 0.016$ (see also Ref.~\cite{Alarcon:2012nr}).  There are much larger discrepancies in the published values for $f_{Tu}$ and $f_{Td}$. In Refs.~\cite{Alarcon:2011zs,Hoferichter:2015dsa}, it is suggested that $f_{Tu} \approx 0.02$ and $f_{Td}\approx 0.04$. The combination that appears in the direct detection amplitude is therefore $f_{Td} -f_{Tu} +  f_{Ts} - \frac{2}{27} f_{TG}\simeq -0.007$ in \micromegas, but one should bear in mind that, because of large cancellations, this is subject to ${\cal O}(1)$ uncertainties.

\Eqref{eq:finalcoupling} displays the $m_h^{-2}$ dependence common to all Higgs-mediated processes, which have cross sections that are currently being explored at direct detection experiments.  At the same time, the $M_1 m_Z/\mu^2$ prefactor signals a further suppression from the Bino-ness of the neutralino dark matter.  This implies that cross sections in this scenario are expected to be significantly smaller than in other models with Higgs-mediated interactions.  It is particularly interesting to see whether these cross sections stay above the neutrino floor, and also how they depend on $|\mu|$, which is often taken as a simple indication of the naturalness of a SUSY model.  To explore these issues, we now turn to a numerical analysis of the direct detection cross section.

\subsection{Spin-Independent Cross Sections}
\label{sec:SI}

We use the package \micromegas~\cite{Belanger:2004yn,Belanger:2014vza} to calculate the particle spectrum and to evaluate the direct detection cross sections.  The 4th-generation squarks add small corrections to the MSSM cross section through box and triangle diagrams that induce couplings of the neutralinos to the gluon content of the nucleons: however, if the squark masses $m_{\tilde q_4}$ are sufficiently larger than $m_{\chi} + m_{q_4}$, where $m_\chi$ is the dark matter mass, then these corrections can be safely ignored.  In this region of parameter space, the MSSM model used by \micromegas\ needs no alteration to accurately estimate the direct detection cross section. 

We determine the SI cross section at several thousand randomly-selected points in parameter space, within the ranges shown in Table~\ref{table:parambounds}.
The 3rd-generation squark mixing is turned off by fixing $A_t - \mu \cot\beta = 0$ and $A_b - \mu\tan\beta=0$, as in the fourth generation.  We fix the gaugino masses to the unification ratios $M_1:M_2:M_3 = 1:2:7$, and we consider the range $200~\gev < M_1 < 700~\gev$.

\begin{table}[tb]
\begin{tabular}{| c | c | c || c | c | c |}\hline
Parameter	& Minimum	& Maximum  	&
Parameter	& Minimum	& Maximum \\ \hline
$M_1$	& 200~\gev	& 700~\gev 	&
$m_{\tilde d},m_{\tilde u}$& 1.2~\tev & 4.0~\tev \\
$\mu$	& $M_1+ 20~\gev$& 12.8~\tev 	&
$m_{\tilde s},m_{\tilde c}$& 1.2~\tev & 4.0~\tev \\
$m_A$	& 0.8~\tev 		& 10~\tev 		&
$m_{\tilde b}$& 0.9~\tev & 4.0~\tev \\
$\tan\beta$& 5			&	 50 		&
$m_{\tilde t}$& 0.9~\tev & 4.0~\tev \\ \hline
\end{tabular}
\caption{List of relevant parameters to the direct detection cross section, and the ranges used for our \micromegas\ calculation.}
\label{table:parambounds}
\end{table}

With these parameters, there is always a choice of 4th-generation parameters that can give the correct thermal relic density.  Since these 4th-generation parameters do not enter the direct detection cross sections, the impact of restricting our models to those with the correct thermal relic density is simply that it restricts the mass range to $200~\gev < M_1 < 540~\gev$ for QUE models, while the entire range $200~\gev < M_1 < 700~\gev$ is accessible for QDEE models. Note that the parameter scan does include values of $m_A$ and $\tan \beta$ for which resonance annihilation effects are important and our calculation of the relic density is not reliable. However, such points only make up a small fraction of the parameter space and we have checked that excluding them does not significantly alter the the direct detection results shown below. 

In \figref{higgsinofrac}, we show the relationship between the \cSI\ cross section and the Bino, Wino, and Higgsino composition of the neutralino dark matter. The cross section predicted by \eqref{eq:finalcoupling} is plotted along with the \micromegas\ results; we see that the analytic approximation is an excellent approximation for many of the models. Smaller values of $|\mu|$, when the Higgsino fractions are largest, correspond to the largest direct detection cross sections.
The width of the bands in \figref{higgsinofrac} is due to the variation of the squark and neutralino masses, and the variation in $\tan\beta$. These effects combine to change the cross section by $\mathcal O(1)$ factors, which are quite small compared to the five orders of magnitude explored by varying $N_{41}^2$. 

\begin{figure}[t]
\includegraphics[width=.48\linewidth]{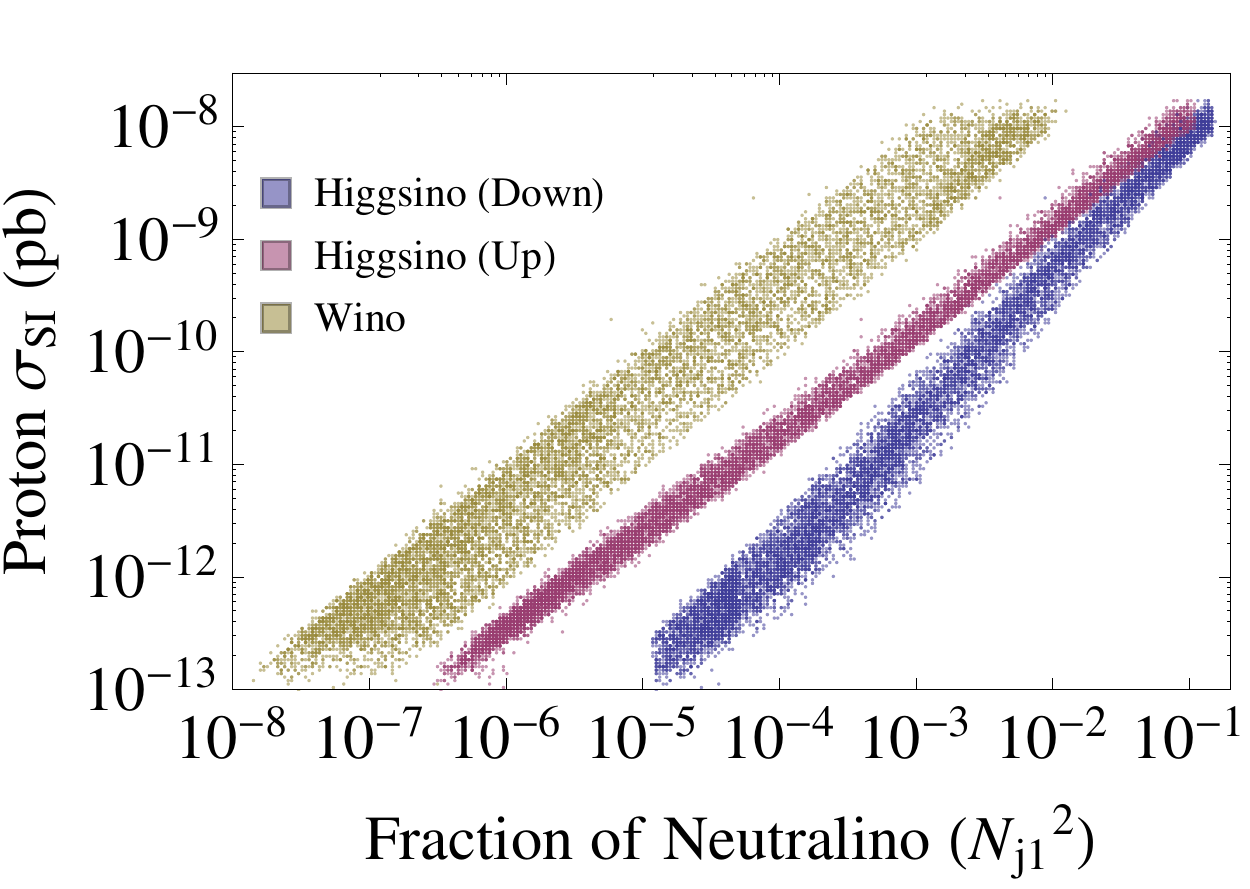} \
\includegraphics[width=.50\linewidth]{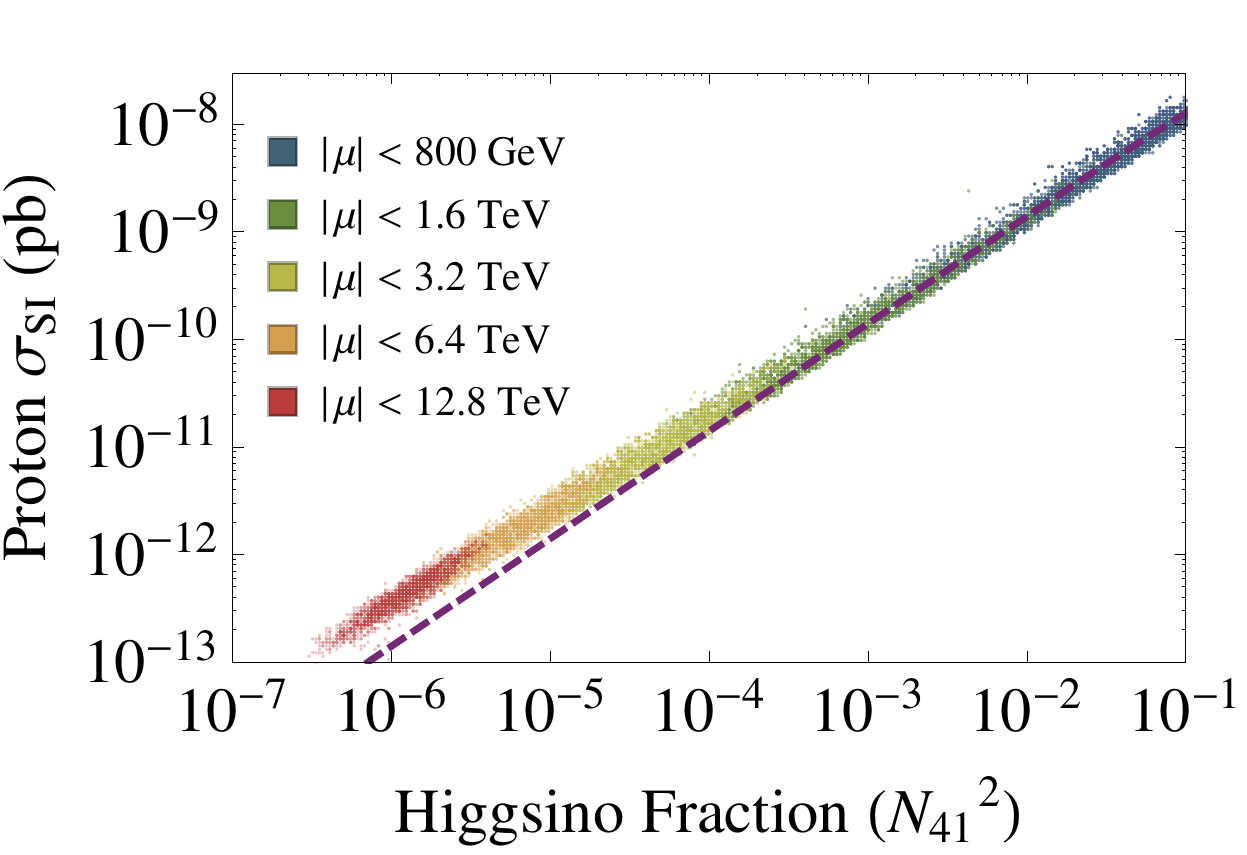}
\vspace*{-0.1in}
\caption{Left: For MSSM4G models, the correlation of the neutralino dark matter's $\tilde{W}$, $\tilde{H}_d$, and $\tilde{H}_u$ fractions with the SI proton scattering cross section $\sigma_{\text{SI}}^{(p)}$.  Right: For MSSM4G models, the correlation of the neutralino dark matter's $\tilde{H}_u$ fraction with $\sigma_{\text{SI}}^{(p)}$, color-coded by the value of $|\mu|$ for each model point. The dashed line represents the analytic approximation for the cross section given in \eqref{eq:finalcoupling}.
In both panels, points in each scatter plot represent QUE and QDEE MSSM4G models that have 125 GeV Higgs bosons, are consistent with all collider bounds, and have the thermal relic density $\OmegaDM h^2 = 0.12\pm0.012$.
\label{fig:higgsinofrac} }
\vspace*{-0.1in}
\end{figure}

\begin{figure}[t]
\includegraphics[width=.98\linewidth]{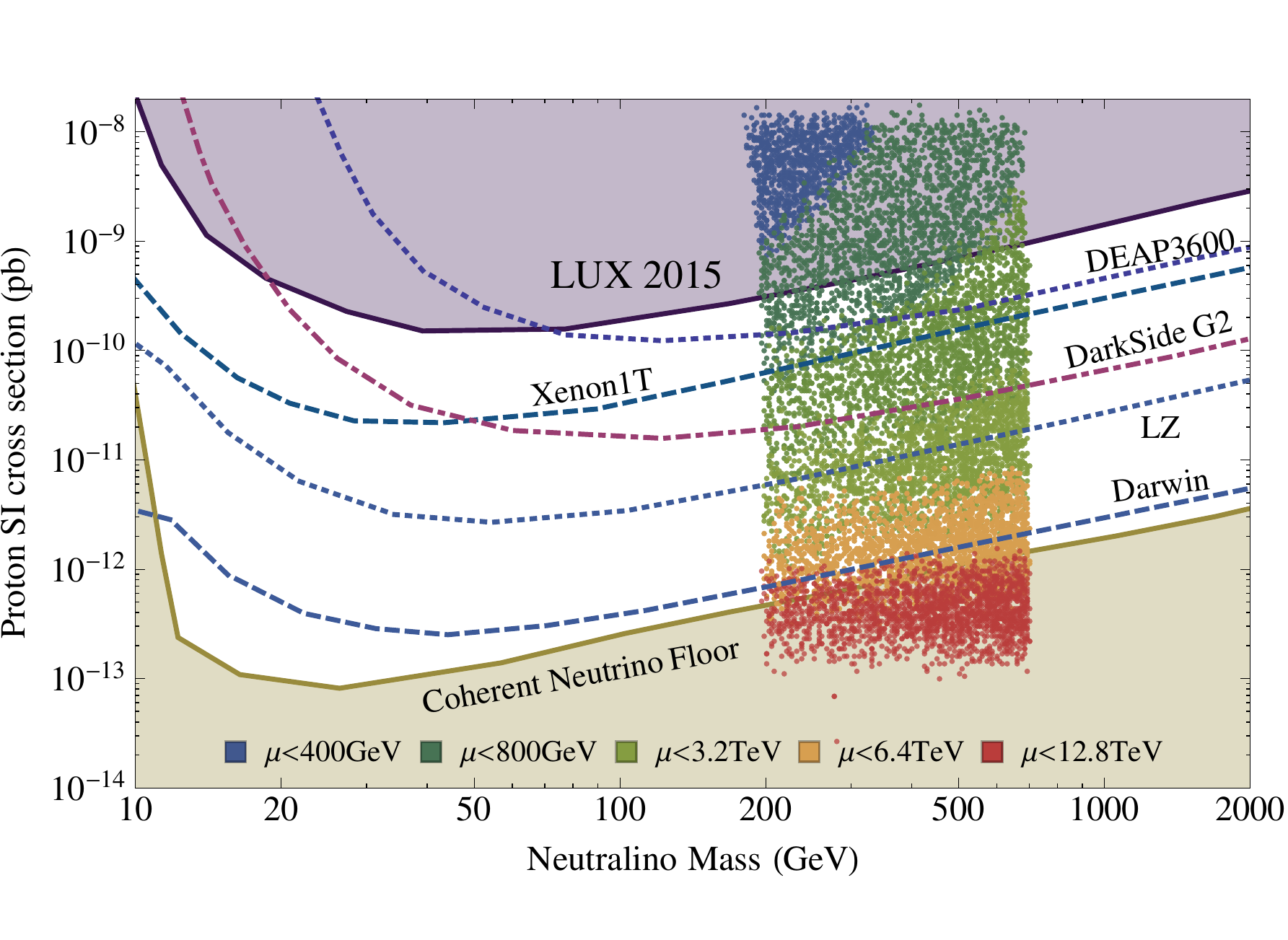} \
\vspace*{-0.15in}
\caption{Scatter plot of theoretical predictions for MSSM4G models in the $(m_{\chi}, \sigma_{\cSI}^{(p)})$ plane. The points represent QUE and QDEE MSSM4G models that have 125 GeV Higgs bosons, are consistent with all collider bounds, and have the correct thermal relic density.  QUE models populate the mass range $200~\gev \alt m_{\chi} \alt 540~\gev$, and QDEE models populate the full range $200~\gev \alt m_{\chi} \alt 700~\gev$.   The points are color-coded by the value of $|\mu|$ in each model point.  The upper shaded region is excluded by the current bound from LUX~\cite{Akerib:2015rjg}, and the dashed contours indicate the projected future sensitivities for DEAP3600~\cite{Amaudruz:2014nsa}, Xenon 1T~\cite{Aprile:2015uzo}, DarkSide G2~\cite{Aalseth:2015mba}, LZ~\cite{Akerib:2015cja}, and Darwin~\cite{Aalbers:2016jon}.  In the lower shaded region, coherent neutrino scattering is a background. 
\label{fig:wideparamscan} }
\vspace*{-0.1in}
\end{figure}

In \figref{wideparamscan} we compare our theoretical predictions to the current experimental bounds from LUX~\cite{Akerib:2015rjg} and the projected 2 ton-year sensitivity of Xenon1T~\cite{Aprile:2015uzo}, as well as several other future experiments.
The current LUX results exclude all of the MSSM4G models generated with $|\mu|<500~\gev$.  For heavier $m_\chi$, models with larger values of $|\mu| \approx 700~\gev$ can be ruled out.  For larger $|\mu|\agt 1\ \tev$ the cross sections are suppressed, as expected, and for $|\mu| \agt 6~\tev$, the cross section drops below the floor from coherent neutrino scattering~\cite{Feng:2014uja}.  Of course, absent a quantitative theory relating the $\mu$-parameter to the SUSY-breaking parameters, such large values of $|\mu|$ require large fine-tuning to obtain the observed weak scale and are typically judged unnatural.  

To summarize, then, for extremely low or high values of $|\mu|$, direct detection cross sections are either excluded or below the neutrino floor, but for a large intermediate region with $500~\gev <|\mu|< 6~\tev$, MSSM4G theories with the correct thermal relic density predict SI scattering cross sections that are not yet excluded, but will be tested by future experiments as they improve their sensitivity down to the neutrino floor.

\subsection{Spin-Dependent Cross Sections}
\label{sec:SD}

Although the SD direct detection cross section is generally larger than the SI cross section, it is much more difficult to probe experimentally, as the \cSD\ cross section does not scale directly with the mass of the nuclei. As a result, current bounds on the neutron \cSD\ cross section are less stringent by a factor of $10^6$.

We use \micromegas\ to predict the proton and neutron \cSD\ cross sections for the same range of models considered in Section~\ref{sec:SI}. As in the SI case, the proton and neutron have similar SD cross sections. It requires different experimental techniques to measure the two cross sections, and several experiments, including PICO-2L~\cite{Amole:2015lsj}, PICO-60~\cite{Amole:2015pla}, and IceCube~\cite{Aartsen:2016exj} probe only the proton SD cross section.

\begin{figure}[t]
\includegraphics[width=.48\linewidth]{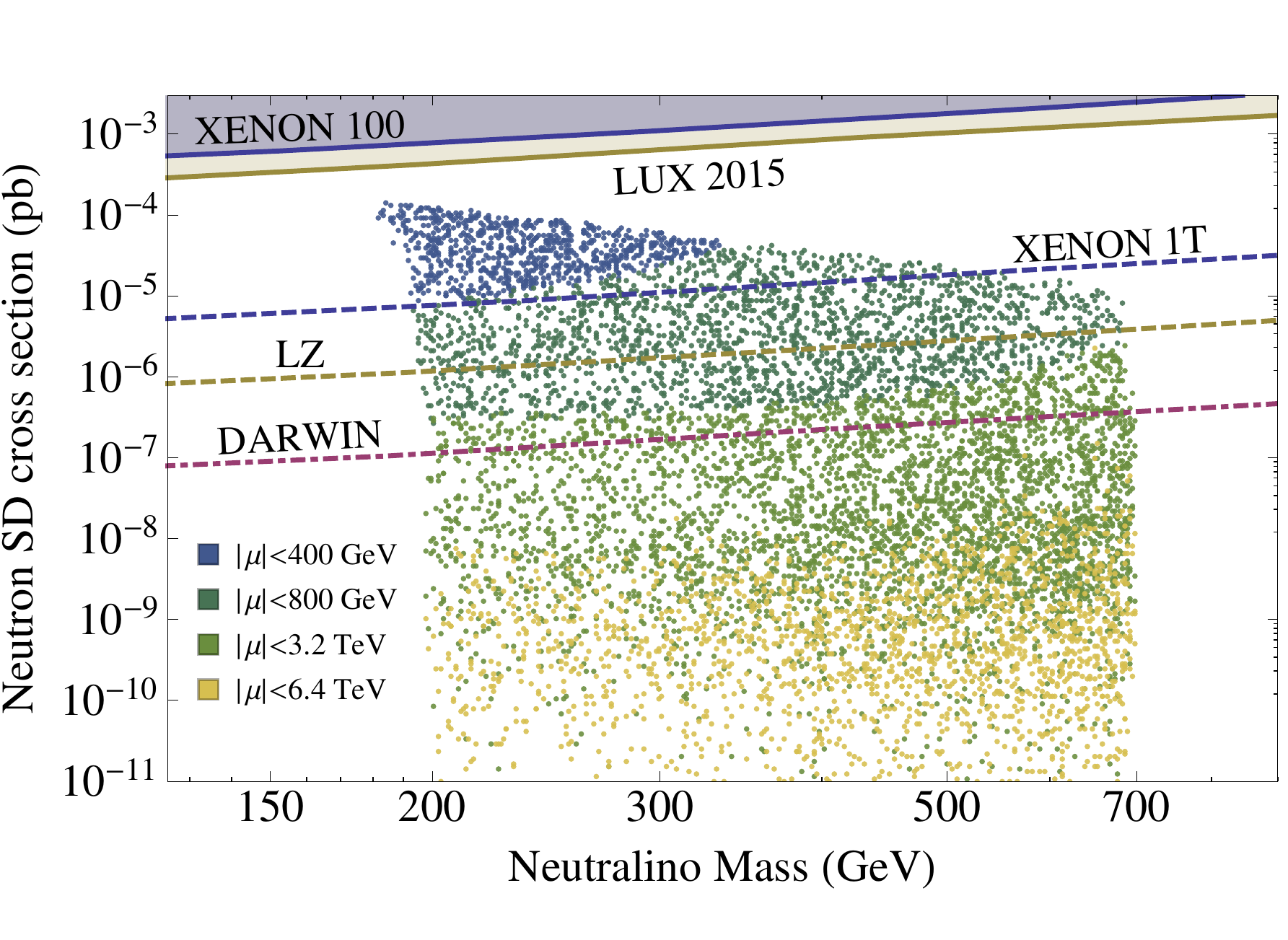}
\includegraphics[width=.48\linewidth]{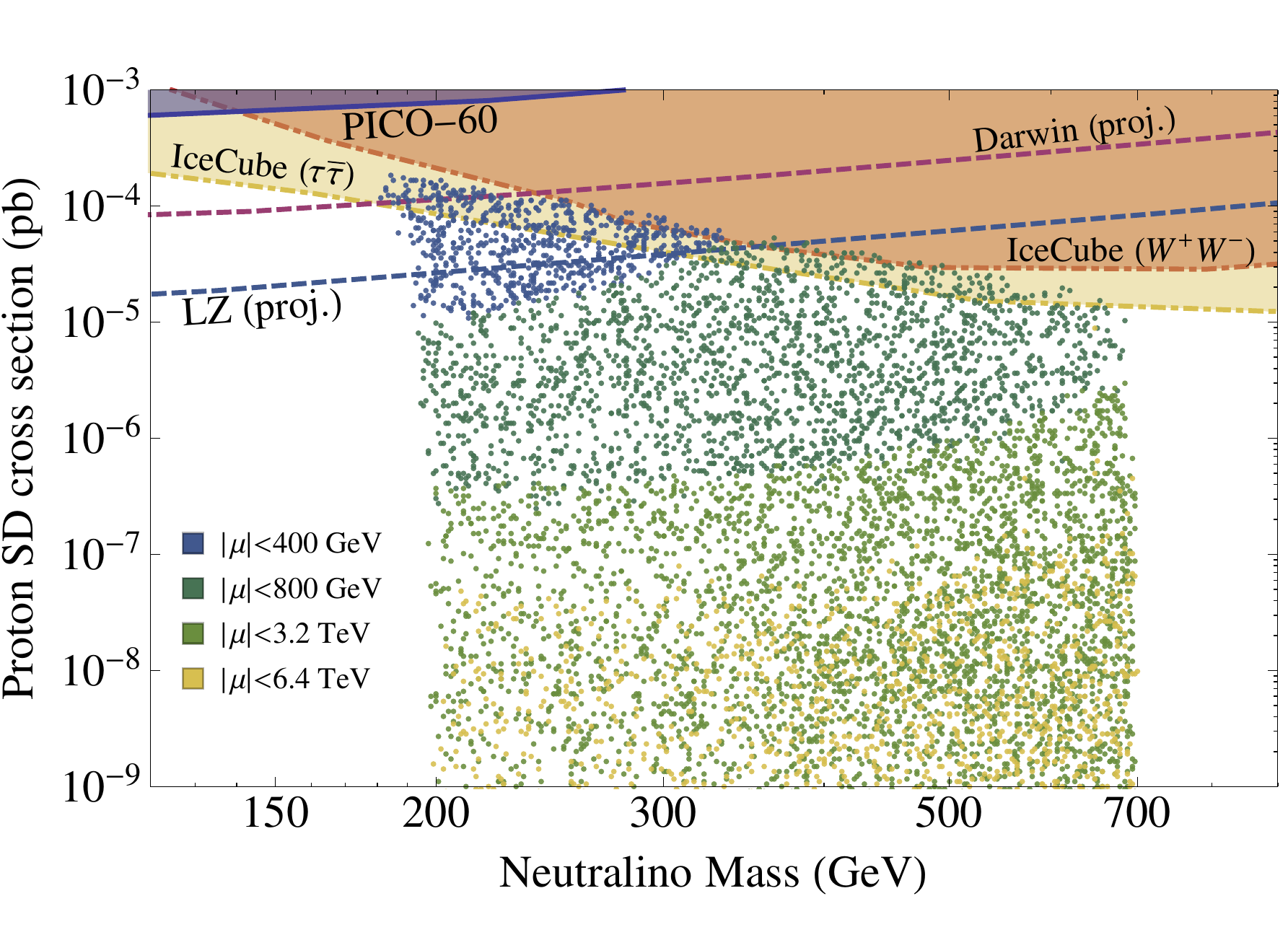} 
\vspace*{-0.1in}
\caption{Left: Predictions for the neutron \cSD\ cross section in MSSM4G models, along with experimental bounds. The shaded regions show the excluded parameter space from Xenon 100~\cite{Aprile:2013doa} and LUX~\cite{Akerib:2016lao}, and the projected sensitivities of LZ~\cite{Akerib:2016lao}, Xenon1T and DARWIN~\cite{Schumann:2015cpa} are given by dashed lines.  Right: Predictions of the proton \cSD\ cross section in MSSM4G models, along with existing bounds from PICO-60~\cite{Amole:2015pla} and IceCube~\cite{Aartsen:2016exj} and the projected sensitivities of LZ and DARWIN. The IceCube bounds assume dark matter pair annihilates to $W^+ W^-$ or $\tau^+ \tau^-$, as indicated.
\label{fig:spindependent} } 
\vspace*{-0.1in}
\end{figure}

In~\figref{spindependent}, the theoretical predictions and experimental bounds are plotted together for the proton and neutron \cSD\ cross sections. The models shown in the two scatter plots are the same set shown in \figref{wideparamscan}, although the models with $\abs{\mu}> 6.4~\tev$ are not shown here.

Of the existing limits from XENON~100~\cite{Aprile:2013doa}, LUX~\cite{Akerib:2016lao}, PICO~\cite{Amole:2015pla}, and IceCube~\cite{Aartsen:2016exj}, only IceCube sets any constraint on the MSSM4G. The limits from IceCube assume that the dark matter annihilates in the Sun to produce either $\tau^+ \tau^-$, $W^+ W^-$, or $b \bar b$. 
In the QUE and QDEE models, Bino annihilation produces taus and $W$ bosons indirectly as decay products of 4th- (or 5th-) generation leptons. As a result, the observed $\tau^{\pm}$ or $W^\pm$ will carry only a fraction of the initial energy, and IceCube becomes somewhat less sensitive to the MSSM4G.  
In \figref{spindependent}, we make the approximation that the energy of the $\tau^{\pm}$ or $W^\pm$ reconstructs only half of the Bino mass. This shifts the published limits from IceCube to higher masses by a factor of two. 

Future experiments such as LZ~\cite{Akerib:2016lao}, Xenon1T and DARWIN~\cite{Schumann:2015cpa} are projected to probe MSSM4G models with $0.4~\tev < |\mu| \lesssim 1~\tev$. However, \figref{wideparamscan} shows that the same experiments will put much more stringent bounds on the SI cross section. Of the models that could be discovered by future \cSD\ experiments, almost all of them have already been ruled out by LUX.  The SI cross section is a much more promising test of MSSM4G models.

\section{Indirect Detection of Dark Matter}
\label{sec:indirect}

One of the primary features of MSSM4G models is that the dark matter has new annihilation channels in the early Universe.  Barring the highly degenerate case where these annihilations are kinematically forbidden in the late Universe, these annihilations then contribute to indirect detection signals. Indeed, the Binos can annihilate to $\tau_{4}$ pairs in the QUE model (and to both $\tau_4$ and $\tau_{5}$ pairs in the QDEE model), which then decay to SM particles. 

The decays of the new leptons arise from the Yukawa mixings with their SM counterparts. These mixings imply decays to $W\nu_{\ell}$, $Z \ell$ or $h \ell$ where $\ell = e,\, \mu, \,$or $ \tau$. It is reasonable to expect that decays to one of the first three generations will dominate, and in this study, we will analyze the special cases where the mixing is purely to one of the three SM lepton generations.
We will label the respective cases as ``$e$-mixing'', ``$\mu$-mixing'', and ``$\tau$-mixing'', after the SM lepton with which $\tau_{4,(5)}$ mixes.

The partial decay widths of vector-like leptons are~\cite{Kumar:2015tna}
\begin{equation}
 \begin{split}
  \Gamma(\tau_{4,5}\rightarrow W\nu_{\ell}) &= 
  \frac{\epsilon^{2}}{32\pi} m_{\tau_{4,5}} r_W (1-r_{W})^{2}(2+1/r_{W})\;,\\
\Gamma(\tau_{4,5}\rightarrow Z\ell) &= 
\frac{\epsilon^{2}}{64\pi} m_{\tau_{4,5}} r_Z (1-r_{Z})^{2}(2+1/r_{Z})\;,\\
\Gamma(\tau_{4,5}\rightarrow h\ell) &= 
\frac{\epsilon^{2}}{64\pi} m_{\tau_{4,5}}(1-r_{h})^{2}\;,
 \end{split}
 \label{eq:vll-decay}
\end{equation}
where $m_{W}$, $m_{Z}$, and $m_{h}$ are the $W$, $Z$, and Higgs boson masses, respectively; $r_{X} = m_{X}^{2}/m_{\tau_{4,5}}^{2}$ for $X = W,Z,h$; $\ell=e,\, \mu, \,\tau$;  and $\epsilon$ parameterizes the mixing between the SM leptons and the new leptons. Note that the $\epsilon$ dependence drops out when calculating the branching ratios. In the limit where $m_{\tau_{4,5}} \gg m_{W},m_{Z},m_{h}$, the branching ratios satisfy $B(W \nu_{\ell}) \! : \! B(Z\ell) \! : \! B(h\ell) = 50\% \! : \! 25\% \! : \! 25\%$, 
which is already almost the case for $m_{\tau_{4,5}} = 200~\gev$.

In the following subsections, we consider the prospects for the indirect detection of dark matter in MSSM4G models through gamma rays, neutrinos, and positrons.

\subsection{Gamma Rays}
\label{subsec:gamma}

Experiments such as Fermi-LAT, H.E.S.S. II, and CTA can search for high-energy photons from the dark matter annihilation in the Galactic Center or in dwarf spheroidal Milky Way satellite galaxies. 

In the $\tau$-mixing case, all the decay products (except for neutrinos) have sizable branching ratios to hadrons, resulting in $\pi^0$ decays that produce a significant excess of gamma rays that may be observed above astrophysical backgrounds.  On the other hand, in the $\mu$-mixing and $e$-mixing cases, although hadronic decays of the $W$, $Z$, and $h$ bosons result in gamma rays, the $\mu$ and $e$ lead to much weaker gamma-ray signals. 

Various experimental collaborations provide current or projected sensitivities to the dark matter annihilation cross section to $W^{+}W^{-}$ or $\tau^{+}\tau^{-}$. We have also analyzed the gamma-ray signal from annihilation to $\mu^{+}\mu^{-}$, but the resulting bounds are very weak and we therefore omit them in this work. 

In the experimental bounds it is assumed that the dark matter annihilates directly to the SM fields, and so their energies are equal to the dark matter mass. In our case the dark matter annihilates to 4th- or 5th-generation leptons, which then decay to SM  fields, resulting in a distribution of final state energies. To test our model against these results we make two assumptions. First, we treat all bosons ($W$,$Z$, and $h$) to be the same and compare the total rate of their production to the limit on the $W^{+}W^{-}$ channel. This is a reasonable approximation since all three have comparable masses and branching ratios to hadrons. Second, we use the average of possible final state energies to compare with the limits. To a good approximation, this average energy is simply $\bar{E} = m_{\tilde{B}}/2$. This is justified by the observation that the energy distribution of the decay products is fairly uniform for non-relativistic mother particles and the fact that the experimental sensitivities are fairly constant as functions of the dark matter mass for the range of masses we are considering. In the following we will consider the sensitivities to the $W^+W^-$ and $\tau^+ \tau^-$ channels separately.  In a more thorough analysis, one would combine these results, resulting in greater sensitivity or more stringent limits. 

Given the smallness of the dark matter velocity in the late Universe, the thermally-averaged cross section is dominated by the $S$-wave piece. The only relevant process is the annihilation to fermions through sfermion exchange which, assuming the left- and right-handed sfermions are degenerate, is given by
\begin{equation}
\langle \sigma v \rangle = \frac{g^{4}_{Y} Y_{L}^{2}Y_{R}^{2}}{32 \pi}\frac{m_{f}^{2}}{m_{\tilde{B}}}
\frac{\sqrt{m_{\tilde{B}}^{2}-m_{f}^{2}}}
{ \left (m_{\tilde{B}}^{2}+m_{\tilde{f}}^{2}-m_{f}^{2}\right )^{2}} \; \; ,
\label{IDxsection}
\end{equation}
where $g_Y \simeq 0.35$ is the U(1)$_Y$ gauge coupling, $Y_{L}$ and $Y_{R}$ are the left and right hypercharges respectively (in the convention where $Q=T_{3}+Y/2$), $m_{f}$ is the fermion mass, $m_{\tilde{f}}$ is the sfermion mass, and $m_{\tilde{B}}$ is the Bino mass. One can see that even the top quark contribution, enhanced by the factor $m_f^2$, is suppressed compared to the $\tau_{4,5}$ contribution by a factor of $(\frac{1}{3}\frac{4}{3})^{2}/(2^{2})^{2} = 1/81$ and we therefore neglect the SM contributions.

For presentation purposes we want to reduce the number of independent masses appearing in \eqref{IDxsection}. To maximize the Bino mass, we set $m_{\tilde{B}} = 1.2 \, m_{\tau_{4,5}}$ so it is close to the fermion mass, but far enough away that the velocity expansion gives accurate results. The sfermion masses $m_{\tilde{\tau}_{4,5}}$ are then constrained by the requirement of correct relic density which is measured to be $\OmegaDM h^{2} = 0.1199\pm0.0022 $~\cite{Ade:2015xua}. In the QDEE model $m_{\tau_{4}} = m_{\tau_{5}}$ and $m_{\tilde{\tau}_{4}} = m_{\tilde{\tau}_{5}}$ are assumed in this work.

\begin{figure}[t]
\includegraphics[width=.49\linewidth]{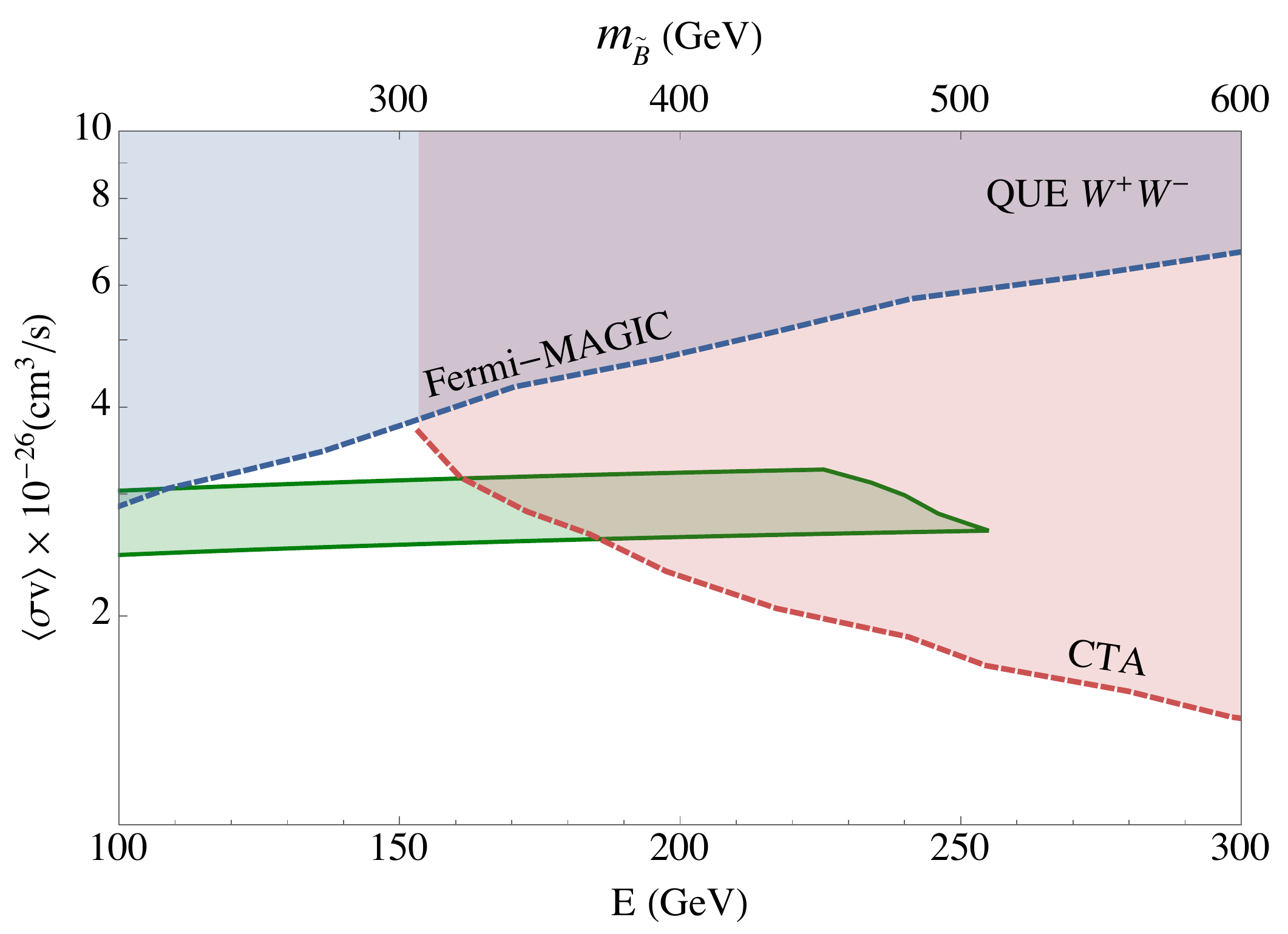} 
\includegraphics[width=.49\linewidth]{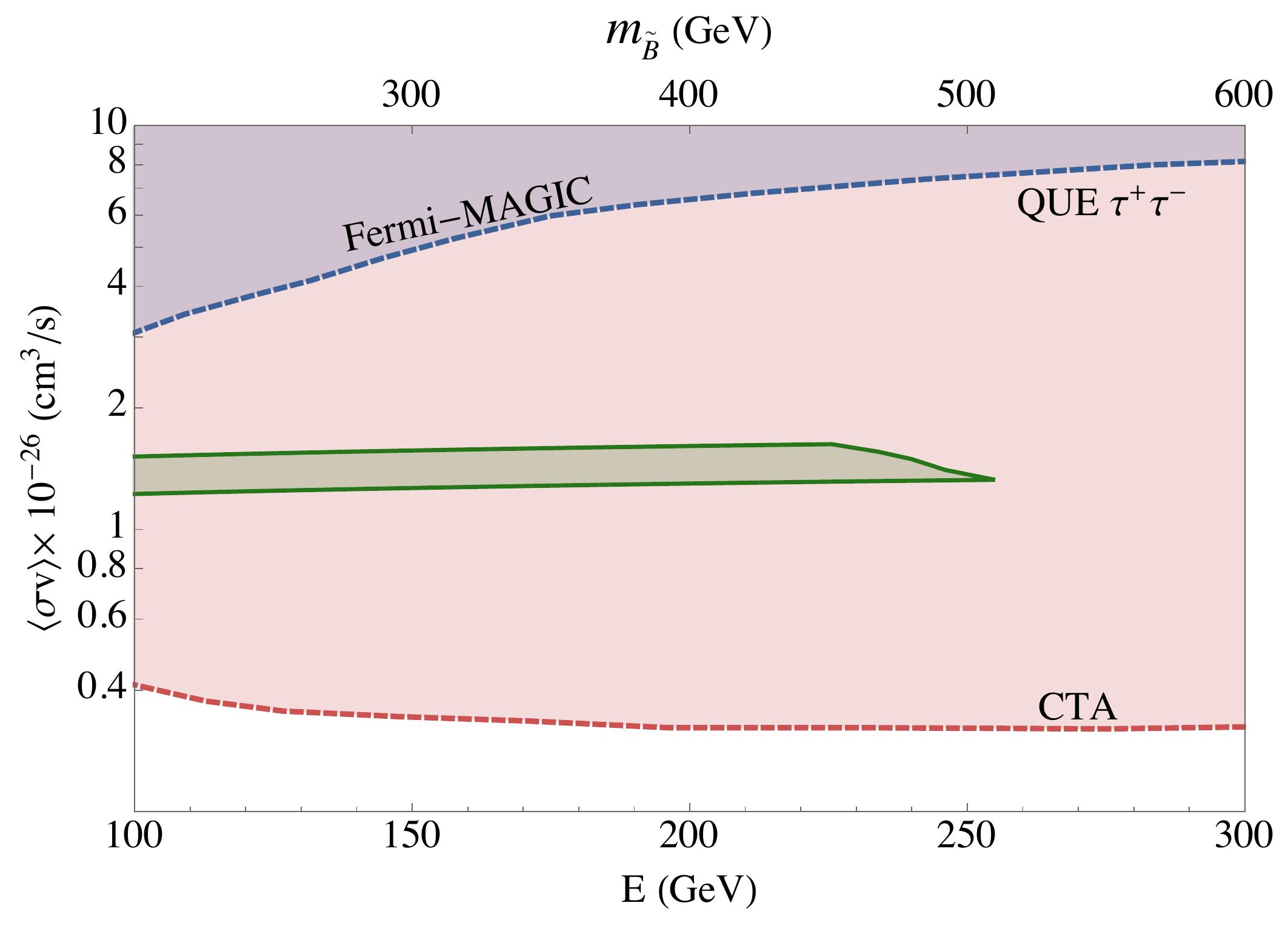} \\
\includegraphics[width=.49\linewidth]{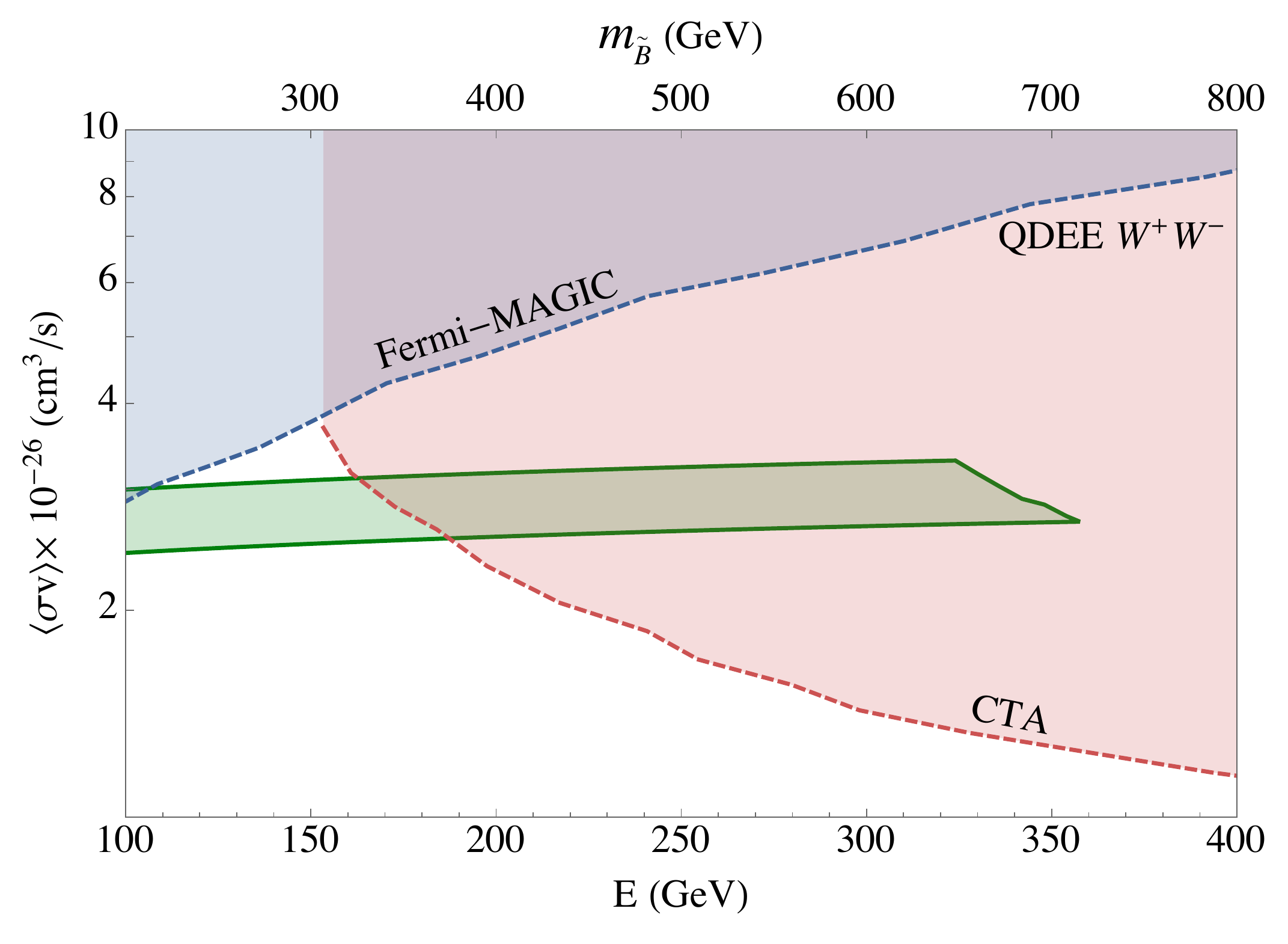} 
\includegraphics[width=.49\linewidth]{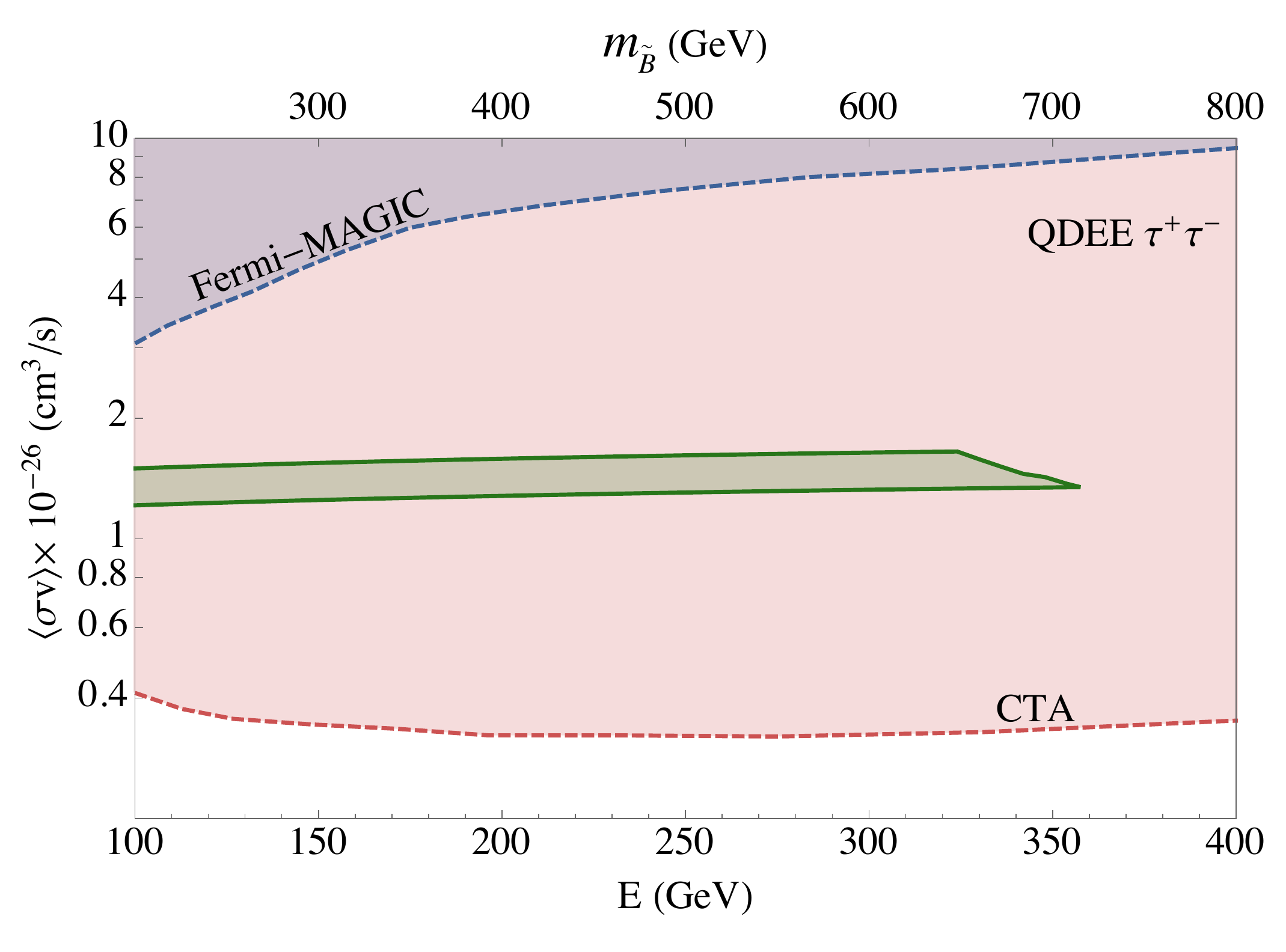} 
\vspace*{-0.1in}
\caption{Theoretical predictions for, and current and future experimental sensitivities to, the annihilation cross sections to $W^{+}W^{-}$ (left) and $\tau^{+} \tau^{-}$ (right) final states in the QUE (top) and QDEE (bottom) MSSM4G models as functions of the dark matter mass (top axis) and average energy $\bar{E} = m_{\tilde{B}}/2$ of the annihilation products (bottom axis). The green-shaded regions are the theoretical predictions for models with thermal relic density in the range $\OmegaDM h^{2} = 0.12 \pm 0.012$; decays to 3rd-generation leptons are assumed for the $\tau^+ \tau^-$ panels.  The dashed blue lines are the existing dwarf bounds from the combined MAGIC and Fermi-LAT data, and the dashed red lines are the CTA projections for Galactic Center sensitivities assuming 500 hours of observation time and an Einasto dark matter profile.  \label{fig:IDQUE} }
\vspace*{-0.1in}
\end{figure}

The theoretical predictions are shown in \figref{IDQUE} for the QUE and QDEE models along with current and future experimental sensitivities. The green strips contain the predictions for MSSM4G models with the correct thermal relic density to 10\%. These strips can be extended to lower masses, although these values are less interesting in light of collider bounds on Binos. On the other hand, extending the strip to higher masses would re-introduce the overclosure problem of Bino dark matter. 

There are two things to note when comparing the theory predictions with the published experimental sensitivities. First, as mentioned above, the energy of our final state particles is roughly half of the dark matter mass, which means the experimental bounds have twice the mass reach. Second, the $\tau_{4,5}$ leptons decay to $h\tau$ and $Z\tau$ only half of the time, which reduces the annihilation cross section limit by a factor of two compared to the more common case where the dark matter annihilates directly to taus.

The strongest current limits come from a combined analysis of MAGIC and Fermi-LAT observations of dwarf spheroidal satellite galaxies \cite{Ahnen:2016qkx}. The limits are barely at the threshold of probing our model and are not expected to improve much in the future. H.E.S.S. II is expected to announce limits that are slightly stronger, but still fairly weak. 

CTA, on the other hand, has the ability to probe a large portion of the parameter space through the $W^+ W^-$ channel and can probe the $\tau$-mixing scenario completely through the $\tau$ channel with 500 hours of observation of the Milky Way Galactic Center ~\cite{Carr:2015hta}. The number of years this will take depends on the fraction of arrays that go online during the first run, which is subject to funding. Optimistically the results shown should be available after less than 3 years of running. The bounds from the $W^+ W^-$ channel, although unable to probe Bino masses below around $340~\gev$, are applicable to all three mixing scenarios. Therefore, in the $e$- and $\mu$-mixing scenarios, this limit needs to be complemented by a different search method.

There are, however, a couple of caveats. First, the limits assume an Einasto dark matter profile; less cuspy profiles give a signal weaker by up to two orders of magnitude. This is mainly due to the uncertainty in the $J$-factors for Galactic Center observations \cite{Abazajian:2015raa}.  We note that the corresponding uncertainty on limits from Fermi-MAGIC observations of dwarf spheroidal galaxies is not as significant \cite{Ackermann:2015zua,Bonnivard:2015xpq}. Second, as we approach the coannihilation domain, the Bino mass needs to be larger to retain the desired thermal relic density.  Since coannihilation does not take place in the late Universe, the indirect detection signal will be weaker, according to \eqref{IDxsection}.

\subsection{Neutrinos and Positrons}
\label{subsec:notgamma}

In principle, it is possible to place limits on indirect detection from IceCube neutrino observations~\cite{Aartsen:2015xej}. In these MSSM4G models, the leading signal is from the decays $\tau_{4,5} \to W \nu$, which produce the most energetic neutrinos.  Softer neutrinos are also produced as secondary decay products.  Unfortunately, the limits on the annihilation cross section from IceCube are larger than $10^{-24} \;\cm^{3}/\s$ and are therefore far less sensitive than gamma-ray searches. 

Dark matter annihilating to positrons is also an important signal, but here the prospects are less clear.   In the $\tau$-mixing scenario, the data can be well fit by assuming dark matter annihilation to $\tau^{+}\tau^{-}$ with cross section $\langle \sigma v \rangle = 6.8^{+1.4}_{-3.3} \times 10^{-24}\;\cm^{3}/\s$~\cite{DiMauro:2015jxa}, which is two orders of magnitude larger than one would expect from a thermal relic annihilating primarily through $S$-wave.  The corresponding cross sections in the $e$- and $\mu$-mixing scenarios are 
$\langle \sigma v \rangle = 5.2^{+1.4}_{-3.8} \times 10^{-27}\;\cm^{3}/\s$ and $\langle \sigma v \rangle = 8.4^{+7.7}_{-3.0} \times 10^{-26}\;\cm^{3}/\s$, respectively, and so much closer to those of thermal relics.  Given the large uncertainties in astrophysical backgrounds, however, it appears that in MSSM4G QUE and QDEE models, the prospects for a compelling indirect detection signal are stronger in gamma rays than in positrons.

\section{Collider Signals}
\label{sec:colliders}

Given thermal relic density constraints, the 4th- and 5th-generation leptons and sleptons in MSSM4G models cannot be arbitrarily heavy.  As a result, MSSM4G models have two robust signatures at hadron colliders: one is Drell--Yan pair production of the 4th- (and 5th-) generation lepton(s) $\tau_{4(,5)}$, and the other is Drell--Yan pair production of their superpartners $\tilde\tau_{4L,4R(,5L,5R)}$, which are the next-to-lightest SUSY particles. With a large mixing parameter $\epsilon$ between the SM and extra-generation lepton(s), we also have single production of $\tau_{4(,5)}$~\cite{Coutinho:1998bu}.

The decays of the extra particles are controlled by the mixing parameter $\epsilon$.
The decay widths of the extra lepton(s) are summarized in \eqref{eq:vll-decay}. The decay length is given by
\begin{equation}
 c\tau\approx\left(\frac{m_{\tau_{4,5}}}{16\pi}\epsilon^2\right)^{-1}
  =\frac{5\times10^{-17}\,\mathrm{m}}{\epsilon^2}
  \cdot\frac{200~\gev}{m_{\tau_{4,5}}}
\end{equation}
for $m_{\tau_{4,5}} \agt 200~\gev$.
The extra sleptons decay through
\begin{equation}
 \tilde\tau_{aM}\to\tau_{a}+\tilde B \ ,
\end{equation}
where $a=4(,5)$ and $M=L,R$, if kinematically allowed.
However, as we will see in \figref{LHC-summary}, this channel is kinematically forbidden in a large portion of the viable parameter regions of MSSM4G obtained in Ref.~\cite{Abdullah:2015zta}, and it is allowed only in a small region of the QDEE models with $m_{\tilde \tau_{4,5}} \sim 450\text{--}500~\gev$.  In the rest of the QDEE parameter space, as well as in all of the QUE parameter space, the sleptons decay through the mixing $\epsilon$ via
\begin{equation}
 \tilde\tau_{aM}\to l_i+\tilde B \ ,
\end{equation}
where $l_i$ is the lepton that mixes with $\tau_a$. This channel gives exactly the same signature as the MSSM right-handed slepton that mixes with the extra sleptons.

Consequently we have three relevant searches for MSSM4G models.
If the mixing is tiny, with $\epsilon \alt 10^{-8}$, searches for long-lived charged particles (LLCPs) are relevant.  With a larger mixing, $\tau_{4(,5)}$ can be searched for by dedicated vector-like lepton searches, and the superpartners by MSSM slepton searches.  With the unified-mass assumptions of \eqsref{eq:unifiedmass-QUE}{eq:unifiedmass-QDEE}, the extra particles in the QUE models are thus equivalent to one vector-like lepton and two right-handed sleptons, while in QDEE models, they are equivalent to two vector-like leptons and four right-handed sleptons.\footnote{Note that the production cross section of $\tilde\tau_{aM}$ is the same as that of the MSSM right-handed sleptons, despite their being labelled with subscripts  `$L$' and `$R$.'}  The discussion below assumes that the extra lepton(s) and their superpartners mix purely with either the 1st-, 2nd-, or 3rd-generation leptons and sleptons, respectively, but it can also be generalized to more complicated mixing patterns.
In fact, LLCP searches are obviously independent of the mixing patterns, and sensitivities of the vector-like lepton and slepton searches would be worse in the case of multiple decay channels.

\subsection{LLCP searches}

%
%

LLCPs are searched for by their anomalous energy loss and longer time-of-flight at the LHC. The CMS Run~1 search excluded leptons with charge $\pm e$ lighter than $574~\gev$, and staus lighter than $340~\gev$, assuming only Drell--Yan pair-production~\cite{Chatrchyan:2013oca,CMS-PAS-EXO-15-010}, and the ATLAS Collaboration provided similar exclusion limits~\cite{ATLAS:2014fka}.

Interpreting this bound under the unified-mass assumptions, one finds that
the QUE models with $m_{\ell_4}<574~\gev$ or $m_{\tilde \ell_4}<410~\gev$ are excluded, while in the QDEE model the regions with $m_{\ell_4}<650~\gev$ or $m_{\tilde \ell_4}<470~\gev$ are excluded, if the relevant particles are effectively stable in collider detectors. Therefore, all the parameter regions of the QUE models, and most of them of the QDEE models, which is summarized in Ref.~\cite{Abdullah:2015zta} (see also \figref{LHC-summary}), are already excluded if $\epsilon \alt 10^{-8}$. The remaining region of the QDEE models with $650~\gev<m_{\ell 4} \alt 700~\gev$ is expected to be covered soon at Run~2 of the LHC~\cite{Feng:2015wqa}.

For slightly larger $\epsilon$, the leptons $\tau_{4(,5)}$ have an intermediate decay length $1\,\text{mm} \alt c\tau \alt 1\,\text{m}$ and their superpartners remain effectively stable at colliders, or both leptons and sleptons can have intermediate decay lengths.
Charged particles with intermediate decay lengths are searched for at the LHC but constrained less severely~\cite{Aad:2015qfa}, while stable $\tilde\tau_{aM}$'s lighter than $\sim 800~\gev$ may be discovered at LHC Run~2 with $300~\ifb$ of data~\cite{Feng:2015wqa}.

\subsection{Vector-like Lepton Searches}

LHC searches for vector-like leptons are performed under the assumption that they mix only with electrons or with muons, which partially excludes the region with $m<200~\gev$~\cite{Aad:2015dha} (see also Refs.~\cite{Falkowski:2013jya,Chatrchyan:2014aea,Dermisek:2014qca}).
Constraints on vector-like leptons mixed with taus are obtained at LEP, which excluded them with masses less than $101~\gev$~\cite{Achard:2001qw}.

The Run~2 prospects for $\tau$-mixed vector-like leptons are studied in Ref.~\cite{Kumar:2015tna}.
Interpreting their results in our scenarios, we find that the $13~\tev$ LHC with $3000~\ifb$ of data may exclude $\tau_{4(,5)}$ leptons lighter than $234~\gev$ ($264~\gev$) in the QUE (QDEE) model with a very optimistic background estimation.
Consequently, $e^+e^-$ colliders are essential to search for $\tau$-mixed vector-like leptons.
Considering the pair-production $e^+e^- \to \tau^+_4 \tau^-_4$, the ILC with $\sqrt s=1~\tev$ will cover the whole parameter region of the QUE models, while the QDEE model, which is viable for $m_{\tau_{4,5}} \alt 700~\gev$, will be fully covered by $\sqrt{s} \agt 1.4~\tev$.
Models with relatively large mixing parameters, roughly $\epsilon \agt 0.01$, may also be searched for through the single production process $e^+e^-\to \tau^\pm_4\tau^\mp$ at smaller collision energies~\cite{DePree:2008st,Djouadi:2016eyy}.

\newcommand{\mET}{\slashed{E}_{\rm T}}
\newcommand{\PT}{P_{\rm T}}
\newcommand{\TODO}[1]{{\color{red}\bf#1}}
The discovery prospects for $e$- and $\mu$-mixed vector-like leptons are considerably brighter than for the $\tau$-mixed case.
We have performed Monte Carlo simulations to determine the future prospects of searches at LHC Run~2 with $\sqrt s=14~\tev$. A thorough description of the analysis is given in Appendix~\ref{sec:LHC}, and the results are summarized in \tableref{vll-massreach}.

\begin{table}[t]\centering
 \caption{\label{table:vll-massreach}
 Future prospects for searches for vector-like leptons at the 14 TeV LHC for three values of integrated luminosity.  The first table is for the QUE models, and the second for the QDEE models.  We consider vector-like leptons with a mass $m_{\ell_4}\ge200~\gev$; the expressions $0^{+250}~\gev$ etc.\ show that the central value of exclusion or discovery limit is below our model points and we may achieve the limit of $250~\gev$ with $1\sigma$ statistical fluctuation. In the dashed entries the upper limit is less than $200~\gev$ even with $1\sigma$ statistical fluctuation. The CL$_s$ method is used for statistical treatment, where the statistical uncertainty and a $20\%$ systematic uncertainty for the background contribution are taken into account, while the theoretical uncertainty on the signal cross section as well as the NLO correction are not considered.  See Appendix~\ref{sec:LHC} for further details.}
 \begin{tabular}{|c|c|c|c|c|}\hline
\multicolumn{2}{|c|}{QUE model} & $300\ifb$           & $1000\ifb$ & $3000\ifb$ \\\hline
95\% CL exclusion  &$e$-mixed   & $240^{+60}~\gev$     & $310^{+50}_{-60}~\gev$ & $350^{+40}_{-40}~\gev$ \\
                   &$\mu$-mixed & $270^{+50}~\gev$     & $330^{+40}_{-60}~\gev$ & $370^{+40}_{-40}~\gev$ \\\hline
$3\sigma$ discovery&$e$-mixed   & $0^{+250}~\gev$      & $250^{+60}_{-40}~\gev$ & $300^{+50}_{-50}~\gev$ \\
                   &$\mu$-mixed & $0^{+280}~\gev$      & $260^{+70}_{-60}~\gev$ & $320^{+50}_{-40}~\gev$ \\\hline
$5\sigma$ discovery&$e$-mixed   & ---                 & $0^{+210}~\gev$        & $220^{+20}_{-20}~\gev$ \\
                   &$\mu$-mixed & ---                 & $0^{+210}~\gev$        & $240^{+20}_{-20}~\gev$ \\\hline

  \end{tabular}

 \vspace{1em}

 \begin{tabular}{|c|c|c|c|c|}\hline
\multicolumn{2}{|c|}{QDEE model}& $300\ifb$           & $1000\ifb$ & $3000\ifb$ \\\hline
95\% CL exclusion  &$e$-mixed   & $350^{+40}_{-50}~\gev$ & $390^{+40}_{-40}~\gev$  & $430^{+40}_{-40}~\gev$\\
                   &$\mu$-mixed & $360^{+40}_{-40}~\gev$ & $400^{+40}_{-40}~\gev$  & $440^{+40}_{-40}~\gev$\\\hline
$3\sigma$ discovery&$e$-mixed   & $290^{+60}_{-70}~\gev$ & $340^{+60}_{-40}~\gev$  & $380^{+50}_{-40}~\gev$\\
                   &$\mu$-mixed & $310^{+60}_{-50}~\gev$ & $360^{+40}_{-30}~\gev$  & $400^{+40}_{-30}~\gev$\\\hline 
$5\sigma$ discovery&$e$-mixed   & $0^{+200}~\gev$        & $260^{+40}_{-50}~\gev$  & $310^{+20}_{-30}~\gev$\\
                   &$\mu$-mixed & $0^{+260}~\gev$        & $280^{+30}_{-30}~\gev$  & $320^{+40}_{-20}~\gev$\\\hline
 \end{tabular}
\end{table}

\subsection{Extra Slepton Searches}

We now consider searches for the 4th- and 5th-generation sleptons.  As stated above, in a small portion of the QDEE parameter region with $200~\gev < m_{\ell_4} < 230~\gev$ and $420~\gev < m_{\tilde\ell_4} < 510~\gev$, the decay $\tilde\tau_{aM}\to\tau_{a}+\tilde B$ is allowed.  As the 4th- and 5th-generation leptons are much lighter than their superpartners in this region, vector-like lepton searches are expected to be more sensitive than extra slepton searches.
We therefore concentrate on other parameter regions in which the signature is
\begin{equation}
 pp \to \tilde \tau^+_{aM}\tilde \tau^-_{aM}\to (l^+\tilde B)(l^-\tilde B) \ ,
\end{equation}
with $l$ being the charged lepton that mixes with $\tau_a$.
This signature is equivalent to pair-production of right-handed slepton pairs $\tilde l^+_R \tilde l^-_R$ in the MSSM, but with a production cross section that is twice (four times) as large in the QUE (QDEE) models.

For the $e$-mixed and $\mu$-mixed cases, we derive the current bound and future sensitivity from studies of slepton ($\tilde e_{R}, \tilde\mu_{R}$) searches, since electrons and muons have a similar acceptance and efficiency at the LHC.  Current bounds have been obtained by the ATLAS and CMS Collaborations at the 8 TeV LHC~\cite{Aad:2014vma,Khachatryan:2014qwa}, and prospects for LHC Run~2 have been discussed in Ref.~\cite{Eckel:2014dza}.
We re-interpret the ATLAS result at the 8 TeV LHC~\cite{Aad:2014vma} and the results in Ref.~\cite{Eckel:2014dza} in the context of our MSSM4G models.

The results are summarized in \figref{slepton}. For the QUE (QDEE) model, the solid (dashed) lines display the exclusion region; the dark-gray (light-gray) region is excluded by the current 8 TeV bounds, and the other three lines corresponds to the expected sensitivity at 14 TeV LHC with integrated luminosities of 300, 1000, and $3000~\ifb$ from left to right. Small dots show the model point we used in the simulation to determine Run~2 prospects, which is performed with exactly the same method as in Ref.~\cite{Eckel:2014dza}, utilizing \software{MadGraph5\_aMC@NLO}~\cite{Alwall:2014hca}, \software{Pythia\,6}~\cite{Pythia6.4} with \software{Pythia--PGS}, and \software{Delphes\,3.2.0}~\cite{deFavereau:2013fsa} with \software{FASTJET}~\cite{Cacciari:2005hq,Cacciari:2011ma}. A systematic uncertainty of $5\%$ as well as statistical uncertainty is taken into account.

\begin{figure}[t]
\includegraphics[width=.48\linewidth]{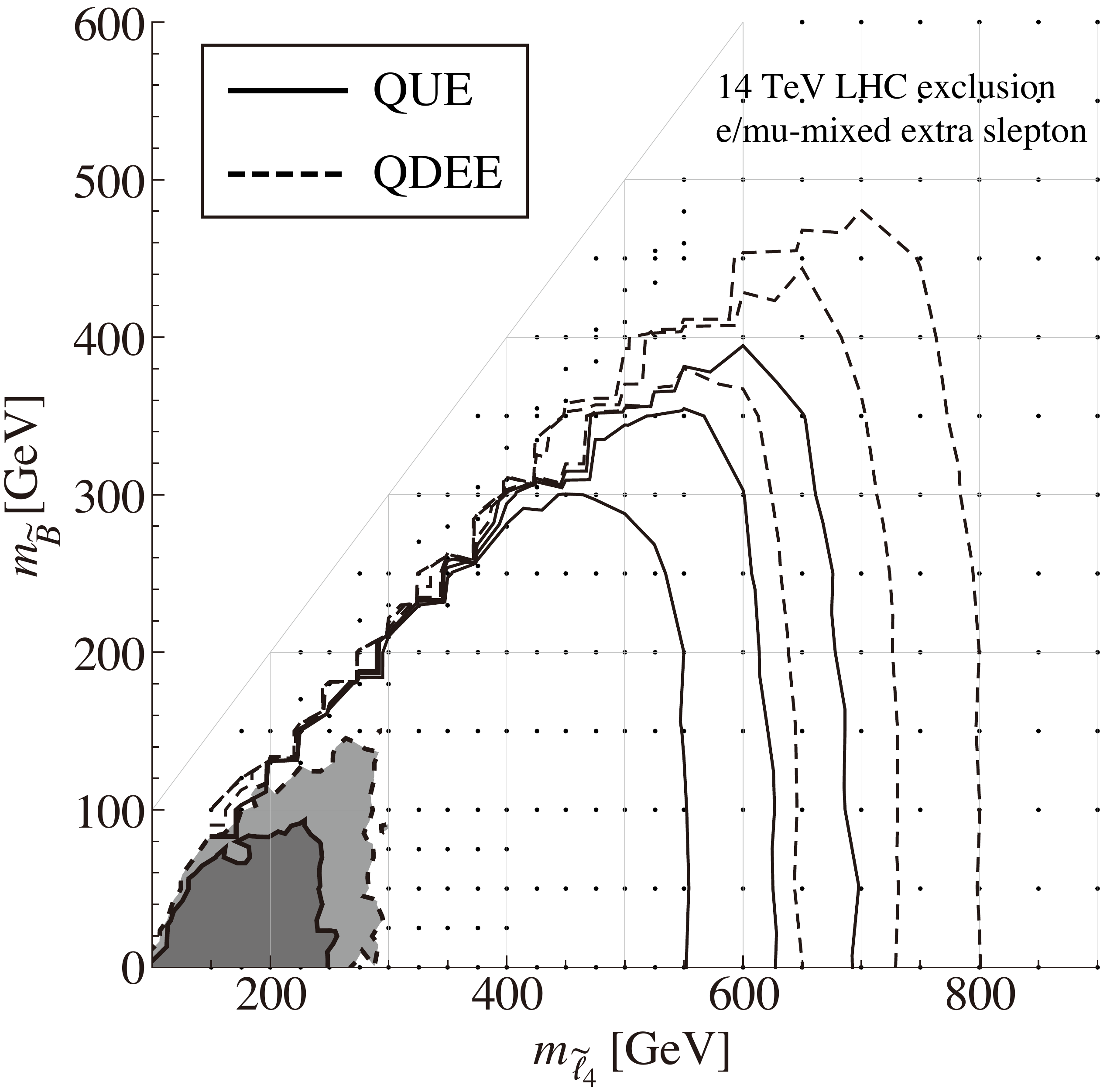}
\vspace*{-0.1in}
 \caption{Current bounds and LHC Run~2 discovery prospects for searches for extra sleptons $\tilde\tau_{4(,5)}$ in MSSM4G models with $e$-mixed or $\mu$-mixed extra lepton generations. For the QUE (QDEE) model, the dark-gray (light-gray) region is excluded by 8 TeV searches~\cite{Aad:2014vma}, and the solid (dashed) contours outline the expected exclusion sensitivities of the 14 TeV LHC with integrated luminosities of 300, 1000, and $3000~\ifb$, from left to right. The small dots show the parameter points we simulated to determine the Run~2 prospects. }
 \label{fig:slepton}
\vspace*{-0.1in}
\end{figure}

For the $\tau$-mixed case, the current bounds on $\tilde\tau_{4,5}$ are no more than $m_{\tilde\ell_4}<120$ (180) GeV~\cite{Aad:2014yka,CMS-PAS-SUS-14-022} in the QUE (QDEE) models, even for $m_{\tilde B}=0~\gev$, which is far below the cosmologically-favored MSSM4G parameter regions.
We have estimated the prospects for searches at LHC Run~2 with two methods.
One method is Monte Carlo simulation.
It is done in a similar way to our analysis of vector-like lepton searches.
As another method, we have rescaled the Run~2 prospects for $e$- and $\mu$-mixed models by the results of the ATLAS Run~1 searches~\cite{Aad:2014yka,Aad:2014vma}.
Both analyses give the result that the 14~TeV LHC is sensitive only below $m_{\tilde B}<210~(140)~\gev$ in the QUE (QDEE) models even with an integrated luminosity of $\int\mathcal L=3000\ifb$.
This region is far below the parameter space motivated by the MSSM4G scenario.
The extra slepton searches are not sensitive to the MSSM4G scenario with mixings with taus, as long as the only available production process is Drell--Yan pair-production $pp\to \tilde \tau^+_{aM}\tilde \tau^-_{aM}$ .

\subsection{Collider Summary and Discussion}

In this section we have discussed the current constraints and future prospects of collider experiments in the MSSM4G models.  Let us interpret the results focusing on the cosmologically-motivated parameter space of the MSSM4G scenario (Fig.~1 of Ref.~\cite{Abdullah:2015zta}).

First, we found that MSSM4G models with $\epsilon\alt10^{-8}$ are mostly excluded by LLCP searches, regardless of the mixing pattern.
A small parameter region of QDEE models with $m_{\ell_4}>650~\gev$ is still valid, and it will be covered in the early stage of the Run~2 LHC.
We also briefly discussed the prospects for models with a slightly larger mixing, $10^{-8}\alt\epsilon\alt10^{-6}$.
Such models will be investigated by searches for leptons decaying inside the detector as well as long-lived sleptons.

With $\epsilon\agt10^{-6}$, the extra particles decay promptly at the LHC, and signatures depend on the pattern of their mixing with SM leptons.
We discussed this case with two assumptions: the mixings are purely with one of the SM three generations, and only the Drell--Yan production of extra particles
($pp\to(Z,\gamma)\to\tau_{a}^+\tau_{a}^-$ and  $pp\to(Z,\gamma)\to\tilde\tau_{aM}^+\tilde\tau_{aM}^-$) are available at the LHC.

For the $\tau$-mixing scenario, i.e., models in which the extra particles mix only with 3rd-generation MSSM leptons and sleptons ($\tau$ and $\tilde \tau$),
we found that the LHC sensitivity is very limited even with $3000\ifb$ data.
The cosmologically favored MSSM4G parameter region requires $m_{\tilde\ell_4}>220\;\gev$, but searches for extra sleptons are expected to be insensitive to this region.
Only a limited region with $m_{\ell_4}<234 \;(264)~\gev$ of the QUE (QDEE) models is expected to be covered by extra lepton searches~\cite{Kumar:2015tna}.
Improvements in tau-tagging techniques may give better, but still limited, sensitivity.
Discovery of the extra leptons as well as exclusion of further region requires $e^+e^-$ colliders, or proton--proton colliders with higher energy.

For models with $e$- or $\mu$-mixing, we found that searches for extra leptons and extra sleptons are both sensitive.
The results of our analyses are summarized in \figref{LHC-summary}, restricting to $m_{\ell_4}>200~\gev$ for simplicity.
The left (right) figure is for $e$-mixing QUE (QDEE) models, and similar results are obtained for $\mu$-mixed models.
In the color-filled regions, one can tune the lepton mass $m_{\ell_4}$ so that the models have a DM relic density $\OmegaDM h^2=0.12$.
The red line in the right figure illustrates $m_{\ell_4}+m_{\tilde B} = m_{\tilde \ell_4}$.
Below this line the extra sleptons decay as $\tilde\tau_{aM}\to\tau_a+\tilde B$.
Our discussion of extra slepton searches is not applicable to this region.
They are valid only above this line, and in all of the (color-filled) region of QUE models, where the extra sleptons decay into $e$ (or $\mu$) and a Bino.

The black lines are the expected exclusion limit at 14 TeV LHC.
Those parallel to the $m_{\ell_4}$-contours are from the extra lepton searches, and the others are from the extra slepton searches.
Dotted, dashed, and solid lines are for integrated luminosities of $\int\mathcal L=300, 1000$, and $3000~\ifb$, respectively.
We found that, in most cases, the extra lepton searches are more sensitive than the extra slepton searches.
This is because the MSSM4G scenario prefers model points at which the extra sleptons and the Bino are rather close in mass.
The degeneracy results in a smaller missing energy from slepton pair-production, and limits the sensitivity of slepton searches.
Even so, it is very interesting that a considerably large portion of the parameter space is expected to be investigated by both of the searches; simultaneous appearance of excesses in both searches will be a very strong evidence of the MSSM4G model.

To summarize, the exclusion limit for models with $e$- or $\mu$-mixing are expected to be $m_{\ell_4}<350~(430)~\gev$ for QUE (QDEE) models at the 14 TeV LHC with $3000\ifb$ data.
Further exploration at collider experiments requires more luminosity, more beam energy, or lepton colliders.
For discovery, the extra lepton searches are promising, and their sensitivity is summarized in \tableref{vll-massreach}.

\begin{figure}[t]
\includegraphics[width=.48\linewidth]{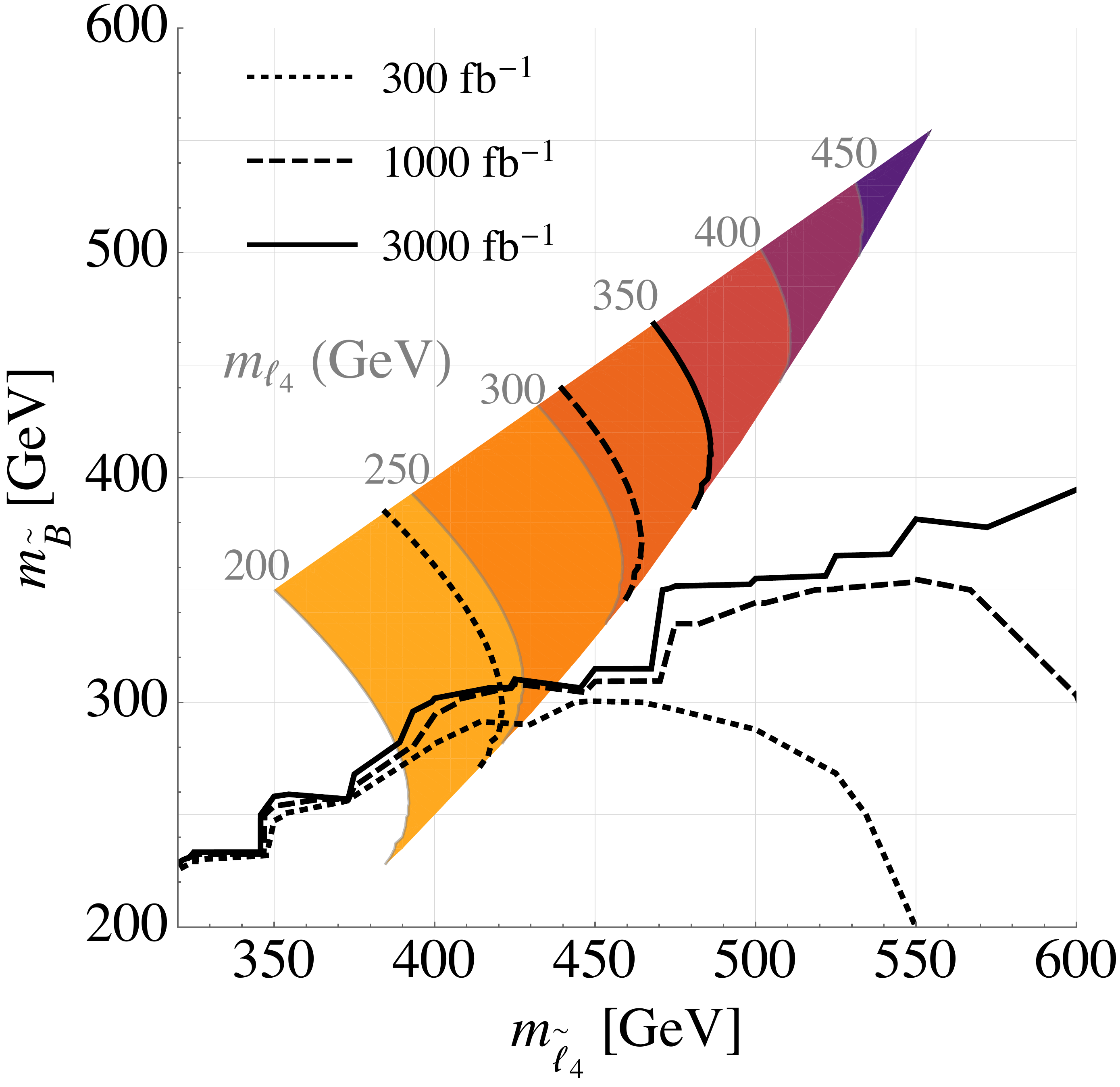} \
\includegraphics[width=.48\linewidth]{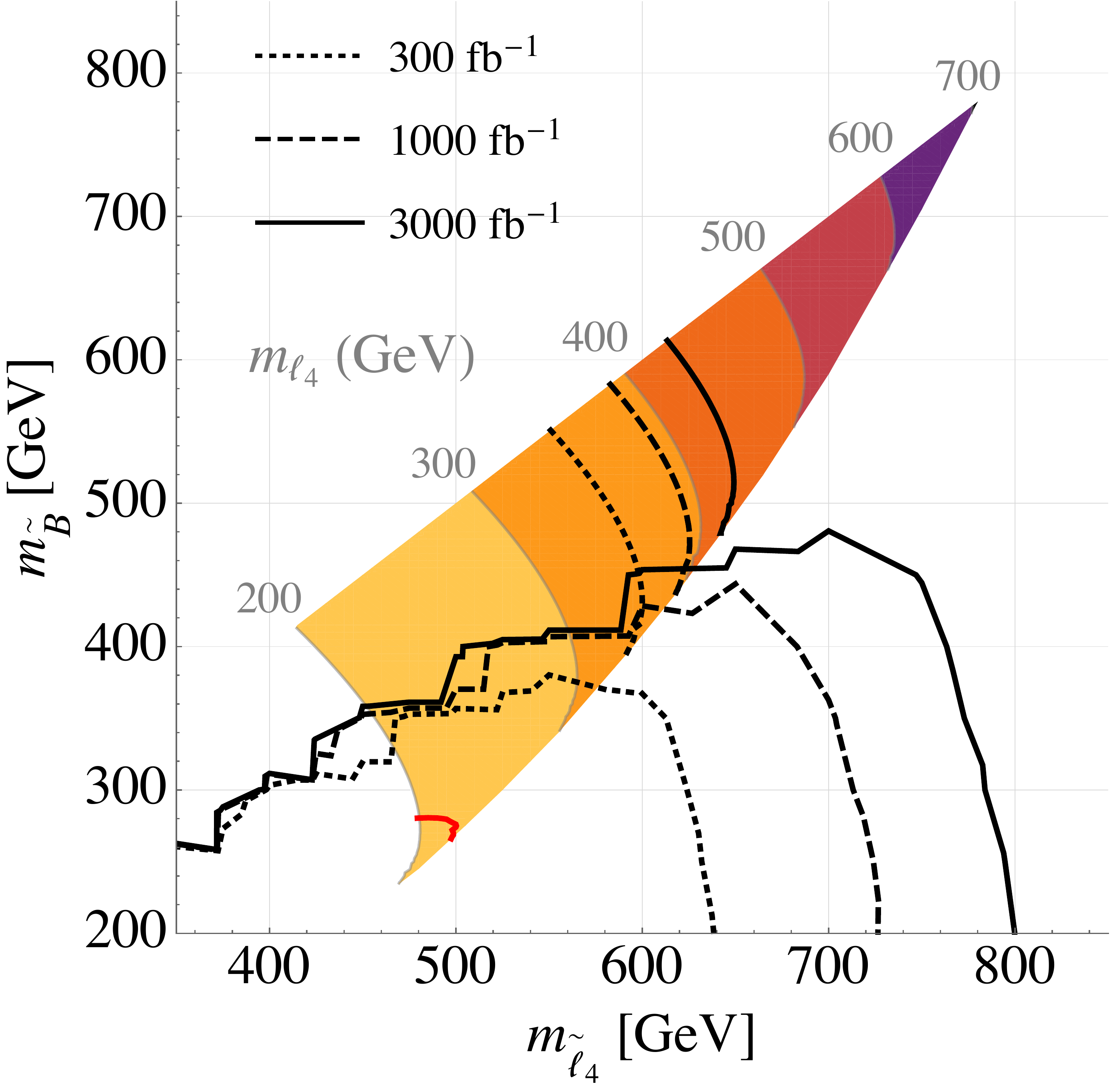}
 \caption{\label{fig:LHC-summary}
 The cosmologically preferred parameter space of QUE (left) and QDEE (right) MSSM4G models, and the exclusion sensitivity of LHC searches in the $e$-mixing case.
  The $\mu$-mixing case results in almost identical sensitivity, while the LHC is expected to be insensitive to the $\tau$-mixing case.
 In both panels, the unified mass relations are assumed and we consider $m_{\ell_4}>200~\gev$. In the shaded regions, $m_{\ell_4}$ is can be tuned so that the model has $\OmegaDM h^2=0.12$; contours of constant $m_{\ell_4}$ are shown in gray.  Outside the shaded regions, the model cannot satisfy $\OmegaDM h^2=0.12$ with $m_{\ell_4}>200~\gev$.
 The black lines are the expected exclusion limits at the 14 TeV LHC.
 Those parallel to the $m_{\ell_4}$-contours are from extra lepton searches.
 The other lines are from extra slepton searches; they are not limited to the color-filled region because they are independent of $m_{\ell_4}$.
 For both searches, dotted, dashed, and solid lines are for an integrated luminosities of $\int\mathcal L=300, 1000$, and $3000~\ifb$, respectively.
 On the red contour in the right plot, the masses satisfy the relation $m_{\ell_4}+m_{\tilde B}=m_{\tilde\ell_4}$.} 
\end{figure}

Let us remark again that this discussion for $\epsilon\agt10^{-6}$ is based on the assumptions that the vector-like lepton(s) has a single dominant mixing and that the other extra particles are not produced.
If other MSSM superparticles are within the reach of the LHC, they will also give some event excess in SUSY searches.
More interestingly, the other vector-like particles are naturally expected to be within the LHC reach.
Extra vector-like quarks are searched for by their characteristic signatures~\cite{Endo:2011xq,Harigaya:2012ir,Endo:2014bsa}, and their superpartners may be found in squark searches.
For models in which the extra vector-like leptons (sleptons) are mixed with more than one generation of SM leptons (MSSM sleptons),
searches for extra leptons are still promising, while those for the extra sleptons suffer from their multiple decay channels.
In general, future prospects for such models are determined by the $e$- or $\mu$-mixed extra lepton searches with the signal yield properly reduced.

\section{Conclusions}
\label{sec:conclusions}

In this work we investigated the current and future prospects of direct, indirect and collider searches for MSSM4G models, where the MSSM is supplemented with vector-like 4th- (and 5th-) generation particles. We began with a brief review of our previous work~\cite{Abdullah:2015zta}, where we showed that such models (specifically the QUE and QDEE models) can enhance the naturalness of the MSSM by increasing the Higgs mass to 125 GeV with relatively light sparticles, preserve gauge coupling unification, and extend the mass of Bino dark matter to the 300--700 GeV range without overclosing the Universe (and without coannihilation).

For direct detection, we found that for neutralino--nucleon scattering, the light Higgs boson-mediated processes dominate over the squark-mediated processes for most of the parameter space, despite the fact that the Higgs-mediated diagram is suppressed by Yukawa couplings and the smallness of the dark matter's Higgsino component. We determined the SI and SD scattering cross sections for various points in MSSM4G parameter space using \micromegas, and for the SI cross section, we derived an accurate analytical expression for the scattering cross section to validate and better understand the results.  

SI searches were found to be much more promising than SD searches.  Current limits from the LUX experiment already exclude all models with $|\mu| < 500\; \gev$, while models up to $|\mu| < 6\; \tev$ will be probed by future planned experiments. Parameter points with larger $|\mu|$ were found to lie below the neutrino floor and would require other approaches, such as directional dark matter detection~\cite{Mayet:2016zxu}. We note, however, that large values of $|\mu|$ are typically considered unnatural and less motivated.  MSSM4G dark matter is therefore an ideal target for current and future direct detection searches. 

To discuss indirect detection and collider searches, we needed to be more concrete about the decay channels of the 4th- (and 5th-) generation leptons. We picked three benchmarks models in which the 4th- (and 5th-) generation leptons have Yukawa mixings with only one of either electrons, muons, or taus, which leads to decays to $W\nu_{l}$, $Zl$, or $h l$ where $l=e,\mu,$ or $\tau$. 

For indirect detection, Bino annihilation to $\tau_{4(,5)}$ followed by decays to SM particles gives a robust gamma-ray signal. Current bounds from the Fermi-MAGIC combined analysis of dwarf spheroidal Milky Way satellites do not yield significant constraints on the MSSM4G parameter space. However, assuming an Einasto (or, in other words, cuspy) dark matter halo profile for the galactic center and 500 hours of observing time, CTA is projected to see a dark matter signal if $m_{\tilde{B}} \agt 340~\gev$ in the $e$- or $\mu$-mixing scenarios, or for the entire range of cosmologically-preferred $m_{\tilde{B}}$ in the $\tau$-mixing scenario.  Prospects for indirect detection through neutrinos at IceCube and through positrons at AMS were found to be significantly less promising.

Finally, we examined the sensitivities of collider searches. In the case of Yukawa mixings of $\epsilon\alt10^{-8}$, the 4th- (and 5th-) generation leptons produced at the LHC are long lived and are either excluded or will be covered by Run~2. The case of $10^{-8}\agt\epsilon\agt10^{-6}$ will also be explored through, for example, displaced vertices. 

For larger mixings, both current and projected bounds depend heavily on the decay products of the new leptons/sleptons. Assuming 3000 $\ifb$ of data at the 14 \tev\; LHC, the $\tau$-mixing scenario will only be probed up lepton masses of $m_{\ell_{4}} < 230\;(250)\;\gev$ in the QUE (QDEE) model. For the $e$- and $\mu$-mixing scenarios the sensitivity reach is up to $m_{\ell_{4}}<350\;(430)\;\gev$ for the QUE (QDEE) model. Interestingly, indirect searches will be sensitive right at the mass threshold where the LHC ceases to be effective, and so the two approaches are highly complementary.

We also analyzed the special regions in parameter space where the decay $\tilde\tau_{4,5}\to\tau_{4,5}+\tilde B$ is allowed. We found that the 14 \tev \; LHC with 3000 $\ifb$ of data will have  poor sensitivity for the $\tau$-mixing case but will fully probe such points for the $e$- and $\mu$-mixing cases.

We have shown that MSSM4G models are perfectly viable on the one hand and predict diverse and promising experimental signals on the other. Although direct detection experiments have strong sensitivities regardless of the details of the extra generation fields, the indirect detection and collider searches are highly dependent on such details and complement each other. Needless to say, all of those projections come with their own caveats. For instance, the direct detection rates are subject to the small uncertainty in the local dark matter density, indirect detection rates are subject to assumptions about halo profiles and our understanding of astrophysical backgrounds, and collider sensitivities depend on the quality of background estimation at higher energies as well as improvements in particle identification techniques.

In summary, MSSM4G QUE and QDEE models are among the motivators of both current and proposed experiments from either the pure (or almost pure) Bino dark matter or the extra generation perspective. It is interesting to continue the search for Bino dark matter with mass $\sim 300\text{--}700~\gev$.  At the same time, we demonstrated both the power and limitations of the upgraded LHC, both important points to take into consideration in discussing proposals for future lepton and hadron colliders.

\section*{Acknowledgments}

The authors thank Jonathan Eckel, Shufang Su, and Huanian Zhang for providing detailed information on their work, and Werner Hofmann, Manoj Kaplinghat, and Stephen Martin for useful discussions.  This work is supported in part by U.S. National Science Foundation Grant Nos. PHY--1316792 and PHY--1620638, Israel Science Foundation Grant No.~720/15, by the United States--Israel Binational Science Foundation Grant No.~2014397, and by the ICORE Program of the Israel Planning and Budgeting committee Grant No.~1937/12.  J.L.F.~was supported in part by a Guggenheim Foundation grant and in part by Simons Investigator Award \#376204.

\appendix

\section{Approximate Analytic Expression for the Spin-Independent Scattering Cross Section of Bino-Like Neutralinos}
\label{sec:ddderivation}

In this Appendix, we derive a simple expression for the differential cross section for SI neutralino--nucleus scattering in the limit where the neutralino is Bino-like.  The resulting expression will require some additional approximations, but will provide an analytic cross-check for the numerical results derived in the body of the paper.  

The SI cross section for neutralinos $\chi$ scattering off a nucleus $N$, with nuclear charge $Z$ and mass number $A$, is~\cite{Drees:1993bu} 
\begin{equation}
\frac{d\sigma}{d\abs{\vec{q}}^2} = \frac{1}{\pi v^2} \left[Z f_p + (A-Z) f_n \right]^2 F^2(Q) \ ,
\label{eq:ddsigma}
\end{equation}
where $\vec{q}$ is the momentum transferred in the interaction; $v$ is the velocity of the dark matter; $f_p$ and $f_n$ are the effective couplings to protons and neutrons, respectively; and $F(Q)$ is the nuclear form factor, where $Q$ is the energy transfer.  

For the form factor, a common parameterization is~\cite{Ahlen:1987mn} 
\begin{equation}
F^2(Q) = e^{-Q/Q_0} \ ,
\end{equation}
where 
\begin{align}
Q_0 &= \frac{1.5}{m_N R_0^2} &
R_0 &=  \left[ 0.3 + 0.91 \left( \frac{m_N}{\gev} \right)^{1/3} \right]\times 10^{-15}\, \text{m} \ .
\end{align}
In the non-relativistic limit, the maximum energy transfer from elastic scattering of dark matter is
\begin{equation}
Q_\text{max} = \frac{2m_N v^2}{\left(1+ m_N/m_\chi\right)^2 } \ , 
\end{equation}
where $m_\chi$ and $m_N$ are the masses of the dark matter and the nucleus, respectively.
For all but the heaviest nuclei, $v^2 m_N^2 R_0^2$ is small enough that $F^2(Q) \approx 1$ is a good approximation.

In the heavy-squark limit, the effective nucleon couplings $f_p$ and $f_n$ are approximately equal and are given by~\cite{Drees:1993bu}
\begin{eqnarray}
\frac{f_{p,n}}{m_{p,n}} &=& 
\sum_{q=u,d,s} \frac{f_{Tq} f_q}{m_q} 
+ \frac{2}{27} f_{TG} \sum_{q=c,b,t} \frac{f_q}{m_q} \ ,
\end{eqnarray}
where $f_{Tq} =  \langle n | m_q \bar q q | n \rangle / m_p$ and $f_{TG}=1 - \sum_{u,d,s} f_{Tq}$. Values for each $f_{Tq}$ are shown in \eqref{eq:scalarquark}.

The neutralino interaction strength is encoded in the parameters
\begin{equation}
f_{q} = \sum_{i=h, H} \frac{g T_{i11} h_{i q q } }{2 m_i^2} 
- \frac{ 1}{4} \sum_{\tilde{q}_j} \frac{X'_{qj1} W'_{qj1} }{m_{\tilde{q}_j}^2 - (m_\chi + m_q)^2 } \ .
\label{direct:effquarkf}
\end{equation}
The first term of \eqref{direct:effquarkf} represents the $t$-channel Higgs exchange diagrams.  The effective Higgs couplings are~\cite{Jungman:1995df} 
\begin{equation}
\begin{split}
 T_{h11} = \sin\alpha \, Q''_{11} + \cos\alpha \, S''_{11} \;,\qquad& 
T_{H11} = -\cos \alpha \, Q''_{11} + \sin\alpha \, S''_{11}\;, \\
Q''_{11} = N_{31} (N_{21} - N_{11} \tan\theta_W ) \;,\qquad&
S''_{11} = N_{41} (N_{21} - N_{11} \tan\theta_W )\;, \\
h_{huu} = -\frac{g m_u \cos\alpha}{2m_W \sin\beta} \;,\qquad&
h_{Huu} = -\frac{g m_u \sin\alpha}{2 m_W \sin\beta}\;, \\
h_{hdd} = + \frac{g m_d \sin\alpha}{2m_W \cos\beta} \;,\qquad&
h_{Hdd} = -\frac{g m_d \cos\alpha}{2 m_W \cos\beta}\;, \\
\sin 2\alpha = - \sin 2\beta \left( \frac{m_{H}^2 + m_{h}^2}{m_{H}^2 - m_{h}^2} \right) ,\qquad&
 \cos 2\alpha = - \cos 2\beta \left( \frac{m_A^2 - m_Z^2}{m_{H}^2 - m_{h}^2} \right).
 \label{eq:ddconstants}
\end{split}
\end{equation}
Here $m_A$ is the CP-odd Higgs masses, $\theta_W$ is the weak mixing angle, and $N_{j1}$ are entries in the matrix $N$ that diagonalizes the neutralino mass matrix, given below in~\eqref{eq:masses}. 

The second term in~\eqref{direct:effquarkf} represents the $s$-channel squark exchange processes.
For the SI amplitude, this requires left--right squark mixings, which we assume are negligible.
In particular, for the third and fourth generations we take them to be zero by tuning $A$-parameters.
As a result, tree-level squark exchange contributes only to the \cSD\ amplitude, and the SI amplitude is dominated by the Higgs-mediated scattering.

In the case where $m_{H} , m_A \gg m_{h}$, we may also neglect the heavy Higgs diagram. In this limit $\alpha \simeq \beta - \pi/2$, so that $\sin\alpha \simeq \cos\beta$ and $\cos\alpha \simeq \sin\beta$.  We consider models with $5 < \tan\beta < 50$, so $\sin\beta\simeq\cos\alpha \simeq 1$ and $\cos\beta\simeq\sin\alpha\simeq0$. With these approximations, 
\begin{align}
T_{h11} \rightarrow N_{41} ( N_{21} - N_{11} \tan\theta_W ) \;, &&
\frac{h_{huu}}{m_u} \rightarrow - \frac{g }{2 m_W} \;, &&
\frac{h_{hdd}}{m_d} \rightarrow \frac{g }{2 m_W} \;.
\end{align}
The effective couplings $f_{p,n}$ can then be expressed very simply as
\begin{equation}
\frac{f_{p,n}}{m_{p,n}} = N_{41} \left[ N_{21} - N_{11} \tan\theta_W \right] \frac{ g^2 }{4m_W m_h^2 }  \left[f_{Td} -f_{Tu} +  f_{Ts} - \frac{2}{27} f_{TG} \right] +  \mathcal O \left( m_{H^0}^{-2} , m_{\tilde{q}}^{-2} \right) .
\end{equation}

To further simplify the expression, we can determine the neutralino mixing matrix factors in terms of the underlying SUSY parameters.  The lightest neutralino $\chi$ can be written in the gauge basis $\{ \tilde{B}, \tilde{W}^3, \tilde{H_d^0}, \tilde{H_u^0} \}$ as 
\begin{equation}
\chi = N^*_{11} \tilde{B} + N^*_{21} \tilde{W}^3 + N^*_{31} \tilde{H}_d^0 + N^*_{41} \tilde{H}_u^0 \;,
\end{equation}
where the matrix $N$ diagonalizes the neutralino mass matrix
\begin{align}
M_\chi =  \left( \begin{array}{c c c c}  M_1 & 0 & - m_Z c_\beta s_W & m_Z  s_\beta s_W \\ 0 & M_2 & m_Z c_\beta c_W & - m_Z s_\beta c_W \\ - m_Z c_\beta s_W & m_Z c_\beta c_W & 0 & - \mu\\ m_Z s_\beta s_W & - m_Z s_\beta c_W & -\mu & 0 \end{array} \right) .
\label{eq:masses}
\end{align}
For $|\mu| > M_1$, the lightest neutralino is primarily Bino with a small Higgsino fraction. Given the gaugino mass unification relation $M_2 = 2 M_1$, the $\tilde{W}$ fraction $\abs{N_{21}}^2$ is negligible compared to the $\tilde{H}$ fractions, as can be seen in \figref{higgsinofrac}.  We may then diagonalize the mass matrix in the limit that the $\tilde{W}$ decouples from the lightest neutralino and $\tan\beta$ is large. In this case we may expand in the small parameter
\begin{equation}
x=\frac{s_W^2 m_Z^2}{\mu^2 - M_1^2 + m_Z ^2 s_W^2} \ ,
\end{equation}
and find that, to leading order in $x$, 
\begin{align}
N_{41} \approx -x \frac{M_1}{m_Z s_W} \ ,
\end{align}
and the neutralino mass is 
\begin{equation}
m_\chi \approx M_1 \left( \frac{\mu^2 - M_1^2 }{\mu^2 - M_1^2 + m_Z^2 \sin^2\theta_W} \right) 
\ .
\label{neutmass}
\end{equation}
For $|\mu| = 250~\gev$ and $M_1 = 200~\gev$, \eqref{neutmass} is accurate to $8\%$. The approximation becomes poorer for smaller values of $\mu^2 - M_1^2$.

The effective neutralino--nucleon couplings $f_{p,n}$ can now be written explicitly in terms of SM and SUSY parameters as
\begin{eqnarray}
\frac{f_{p,n}}{m_{p,n}} &=& \frac{M_1 x}{m_Z \cos\theta_W} \left( \frac{g^2}{4 m_W m_h^2 } \right) \left[ f_{Td} -f_{Tu} +  f_{Ts}- \frac{2}{27} f_{TG} \right] + \mathcal O(x^{2}, m_H^{-2}, m_{\tilde q}^{-2} ) \nonumber\\
&\approx& \frac{M_1 m_Z \tan\theta_W \sin\theta_W}{\mu^2 - M_1^2 + m_Z^2 \sin^2\theta_W } \left( \frac{g^2}{4 m_W m_h^2} \right) \left[ f_{Td} -f_{Tu} +  f_{Ts} - \frac{2}{27} f_{TG} \right] . \label{eq:effcoupling} 
\end{eqnarray}
\Eqref{eq:effcoupling} provides a simple analytic expression for the effective scalar neutralino--nucleon couplings when the squarks are effectively decoupled, $m_A \gg m_{h^0}$, $\tan \beta$ is moderate or large, and the neutralino dark matter is Bino-like.

\section{Monte Carlo simulation of vector-like leptons at the LHC}
\label{sec:LHC}

This Appendix describes our Monte Carlo simulation of searches for vector-like leptons at the 14 TeV LHC.  We focus on vector-like leptons that mix with electrons or muons; the Run~2 prospects for $\tau$-mixed vector-like leptons are studied in Ref.~\cite{Kumar:2015tna}.

\subsection{Analysis procedure}

SM background events are estimated with the Snowmass background set for $14~\tev$ $pp$ colliders~\cite{Anderson:2013kxz,Avetisyan:2013onh,Avetisyan:2013dta}.
Signal events are generated with the same procedure that generated the background, i.e., the hard processes are calculated with \software{MadGraph5\_aMC@NLO}~\cite{Alwall:2014hca}, showering and hadronization are performed with \software{Pythia\,6}~\cite{Pythia6.4} with the \software{Pythia--PGS} interface, and the detector is simulated with \software{Delphes} tuned by the Snowmass Collaboration based on \software{Delphes\,3.0.9}~\cite{deFavereau:2013fsa}, with \software{FASTJET}~\cite{Cacciari:2005hq,Cacciari:2011ma} utilized for jet reconstruction.
In the detector simulation, electrons, muons, and jets are reconstructed and identified based on the same procedure and efficiency as the Snowmass background set. The lepton identification efficiency is 98\%  (99\%) for electrons (muons) with $\PT>10~\gev$ and $|\eta|<1.5$, and jets are reconstructed by the anti-$k_{\rm T}$ algorithm\cite{Cacciari:2008gp} with $R=0.5$. The objects are required to be separated from each other by the procedures in Ref.~\cite{Aad:2014nua}, and electrons and muons forming same-flavor opposite-sign (SFOS) pairs with $m_{\rm SFOS}<12~\gev$ are removed.

We do not include further efficiency factors for lepton identification, reconstruction and isolation, even though the results of our analysis, which focuses on events with multi-leptons, are sensitive to these efficiencies.  This is because these efficiencies are determined only through LHC Run~2 data.  In view of this limitation, the production cross section of the leptons are calculated at tree-level without an NLO $K$-factor, and we refrain from using tau-tagging (therefore taus are classified as jets), despite the fact that taus from $Z$ and $W$ would increase the sensitivity of the searches.  For the same reason $b$-tagging is not utilized; as we will see later, the background from top quark events is small.

Background events from the Snowmass background set and signal events after the \software{Delphes} simulations are then analyzed as follows.
Electrons (muons) with $\PT>20~\gev$ and $|\eta|<2.47$ (2.4) are tagged as ``signal'' electrons (muons), which together we call ``signal'' leptons,\footnote{In this Appendix, $\ell$ denotes electrons and muons, but not taus.} and jets with $\PT>20~\gev$ and $|\eta|<2.5$ are tagged as ``signal'' jets. These objects are used in the analysis described below.

Events with $N_{\ell}\ge 3$ are selected, where $N_{\ell}$ is the number of signal leptons.
The leading (sub-leading) lepton must have $\PT>120~\gev$ ($\PT>60~\gev$).
We define five categories, as described in \tableref{vll-SRs}. Each category is then divided into several signal regions (SRs) as follows:
\begin{itemize}
 \item The $WZ(j)$ category is designed for the signature $\tau^+_4\tau^-_4\to(W\nu)(Z\ell)\to(jj\nu)(\ell\ell\ell)$. This category is divided into two SRs according to the number of $Z$-like lepton pairs $N_{Z(\ell\ell)}$, where a lepton pair is tagged as $Z$-like if it is SFOS and $|m_{\ell\ell}-m_Z|<10~\gev$:
       \begin{itemize}
        \item $WZ(j)^{-}$ for $N_{Z(\ell\ell)}=0$,
        \item $WZ(j)^{Z}$ for $N_{Z(\ell\ell)}\ge 1$.
       \end{itemize}
 \item The $WZ(\ell)$ category is designed for the signature $\tau^+_4\tau^-_4\to(W\nu)(Z\ell)\to(\ell\nu\nu)(\ell\ell\ell)$.
       Two SRs are defined by $N_{Z(\ell\ell)}$, but here a $Z$-like lepton pair must not contain any of the leading two leptons:
       \begin{itemize}
        \item $WZ(\ell)^{-}$ for $N_{Z(\ell\ell)}=0$,
        \item $WZ(\ell)^{Z}$ for $N_{Z(\ell\ell)}\ge 1$.
       \end{itemize}
 \item The $ZZ(j)$ category focuses on the signature $\tau^+_4\tau^-_4\to(Z\ell)(Z\ell)\to(jj\ell)(\ell\ell\ell)$.
       For this category, three flags are defined: $J$ if the event has a jet pair with $|m_{jj}-m_Z|<10~\gev$, $L$ if it has $Z$-like lepton pairs not containing the leading lepton, and $Z$ if the leading lepton does not make a $Z$-like lepton pair with another lepton. Eight SRs are defined according to whether the flags are on or off. For example, $ZZ(j)^{JLZ}$ requires all the flags be on, $ZZ(j)^{Z}$ requires only the $Z$ flag, and $ZZ(j)^{0}$ requires that all the flags are off.
 \item The $ZZ(\ell)$ category is for $\tau^+_4\tau^-_4\to(Z\ell)(Z\ell)\to(\ell\ell\ell)(\ell\ell\ell)$. Three inclusive SRs are defined according to the number of jets: $ZZ(\ell)$ for any number of jets, $ZZ(\ell)^{<2}$ for $N_j<2$, and $ZZ(\ell)^{<1}$ for $N_j < 1$.
\end{itemize}

\begin{table}[t]
\centering
 \caption{\label{table:vll-SRs}
 Definition of signal region (SR) categories. Each category is further divided into SRs, as described in the text.  $N_\ell$ and $N_j$ are the number of signal leptons and signal jets, respectively, and $m_{jj}$ is the invariant mass of the two leading jets. $N_{Z(\ell\ell)}$ is the number of SFOS lepton pairs with $|m_{\ell\ell}-m_Z|<10~\gev$.
 }
 \begin{tabular}{|c|c|c|c|c|c|}\hline
                      & $WZ(j)$    &  $WZ(\ell)$   & $ZZ(j)$       & $ZZ(\ell)$ \\\hline
 $N_\ell$             & $\ge 3$    &  $\ge 4$      & $\ge 4$       & $\ge 5$ \\
 $N_j$                & $\ge 2$    &  $<2$         & $\ge 2$       & ---     \\
 $|m_{jj}-m_W|$       & $<20~\gev$  &  ---          & ---           & ---     \\
 $|m_{jj}-m_Z|$       &     ---    &  ---          & $<40~\gev$     & ---     \\
 $\mET$               & $> 60~\gev$ &  $>100~\gev$   & ---           & ---     \\
 $N_{Z(\ell\ell)}$    &    ---     &  ---          & $\ge 1$       & $\ge 1$ \\\hline
 \end{tabular}
\end{table}

\begin{table}[p]\centering
 \caption{\label{table:vll-BKG}Selection flow of the background events in the vector-like lepton search. Upper bounds on the number of events in each SR, $N_{\rm UL}$, are shown for three values of integrated luminosity, where systematic uncertainty of 20\% as well as statistical uncertainty is included.}
 \begin{tabular}{|l|c|c|c|c|c|c|c|}\hline
                & \multicolumn{4}{|c|}{background cross section [fb]}& \multicolumn{3}{|c|}{$N_{\rm UL}$} \\\hline
                & di-boson & tri-boson & top   & total & $300\ifb$&$1000\ifb$& $3000\ifb$ \\\hline
  $N_\ell\ge 3$ & 222      &  5.1    & 13.4    &   249 & ---      & ---      & ---   \\\hline
  $WZ(j)^-$     & 0.071    & 0.013   & 0.082   & 0.166 &  25.1    &  70.4    & 200   \\
  $WZ(j)^Z$     & 0.643    & 0.071   & 0.183   & 0.898 & 111      & 359      &1060   \\\hline
  $WZ(\ell)^-$  & 0.014    & 0.025   & 0.017   & 0.056 &  11.9    &  27.4    &  71.1 \\
  $WZ(\ell)^Z$  & $<0.001$ & 0.005   & 0.003   & 0.008 &   5.1    &   7.9    &  14.5 \\\hline
  $ZZ(j)^0$     & 0.194    & 0.016   & 0.058   & 0.268 &  37.2    & 111      & 321   \\
  $ZZ(j)^{J}$   & 0.064    & 0.007   & 0.022   & 0.093 &  16.4    &  41.8    & 114   \\
  $ZZ(j)^{L}$   & 0.182    & 0.012   & 0.024   & 0.218 &  31.2    &  91.7    & 263   \\
  $ZZ(j)^{Z}$   & 0.020    & 0.004   & 0.019   & 0.043 &  10.2    &  22.2    &  55.7 \\
  $ZZ(j)^{JL}$  & 0.060    & 0.005   & 0.009   & 0.075 &  14.2    &  35.3    &  94.3 \\
  $ZZ(j)^{JZ}$  & 0.008    & 0.001   & 0.008   & 0.017 &   6.7    &  11.9    &  25.6 \\
  $ZZ(j)^{LZ}$  & 0.020    & 0.004   & 0.019   & 0.043 &  10.2    &  22.2    &  55.9 \\
  $ZZ(j)^{JLZ}$ & 0.008    & 0.001   & 0.008   & 0.017 &   6.7    &  11.9    &  25.5 \\\hline
  $ZZ(\ell)$     &$<0.001$ & 0.005   &$<0.001$ & 0.005 &   4.7    &   6.8    &  11.5 \\
  $ZZ(\ell)^{<2}$&$<0.001$ & 0.003   &$<0.001$ & 0.004 &   4.2    &   5.8    &   9.2 \\
  $ZZ(\ell)^{<1}$&$<0.001$ & 0.001   &$<0.001$ & 0.001 &   3.6    &   4.5    &   6.3 \\\hline
 \end{tabular}
\end{table}

\begin{table}[p]\centering
 \caption{\label{table:vll-signal}Selection flow of the signal events in searches for the $e$- or $\mu$-mixed $\tau_4$ in the QUE model, displayed as a signal cross section in fb. SRs marked with $*$, $\dagger$ and $\ddagger$ are the most sensitive for exclusion at $\mathcal L=300$, 1000, and $3000\ifb$, respectively.}
 \newcommand{\xA}{${}^*$}
 \newcommand{\xB}{${}^\dagger$}
 \newcommand{\xC}{${}^\ddagger$}
 \begin{tabular}{|c|c|c|c|c|c|c|}\hline
  $m_{\tau}$ [GeV], mixing
      & 200, $e$ & 200, $\mu$
      & 300, $e$ & 300, $\mu$
      & 400, $e$ & 400, $\mu$\\\hline
  total         & 95.7  &  96.0  & 21.2  & 21.2  & 6.76   & 6.74   \\\hline
  $N_\ell\ge 3$ & 2.23  &  2.42  & 0.634 & 0.671 & 0.231  & 0.230  \\\hline
  $WZ(j)^-$     & 0.018 & 0.022  & 0.020 & 0.024 & 0.011  & 0.012  \\
  $WZ(j)^Z$     & 0.049 & 0.063  & 0.034 & 0.036 & 0.014  & 0.014  \\\hline
  $WZ(\ell)^Z$  & 0.012 & 0.014  & 0.008\xC & 0.008 & 0.003  & 0.004\xC   \\\hline
  $ZZ(j)^0$     & 0.066 & 0.065  & 0.035 & 0.044 & 0.015  & 0.015  \\
  $ZZ(j)^{J}$   & 0.035 & 0.033  & 0.018 & 0.023 & 0.008  & 0.007  \\
  $ZZ(j)^{L}$   & 0.045 & 0.048  & 0.026 & 0.031 & 0.011  & 0.012  \\
  $ZZ(j)^{Z}$   & 0.039\xA & 0.042\xA  & 0.025\xA\xB & 0.029\xB & 0.010\xA  & 0.012\xB  \\
  $ZZ(j)^{JL}$  & 0.025 & 0.025  & 0.013 & 0.016 & 0.006  & 0.006  \\
  $ZZ(j)^{JZ}$  & 0.021 & 0.022  & 0.013 & 0.015\xC & 0.005  & 0.006  \\
  $ZZ(j)^{LZ}$  & 0.039 & 0.042  & 0.025 & 0.029\xA & 0.010\xB  & 0.012\xA  \\
  $ZZ(j)^{JLZ}$ & 0.021 & 0.022  & 0.013 & 0.015 & 0.005  & 0.006  \\\hline
  $ZZ(\ell)$    & 0.015\xB\xC & 0.014\xB\xC  & 0.005 & 0.007 & 0.003\xC  & 0.002  \\
  $ZZ(\ell)^{<2}$&0.010 & 0.009  & 0.003 & 0.004 & 0.002 & 0.001  \\
  $ZZ(\ell)^{<1}$&0.004 & 0.003  & 0.001 & 0.002 & $8\times10^{-4}$& $6\times10^{-4}$\\\hline
  \end{tabular}
\end{table}

\begin{figure}[t]
\includegraphics[width=.48\linewidth]{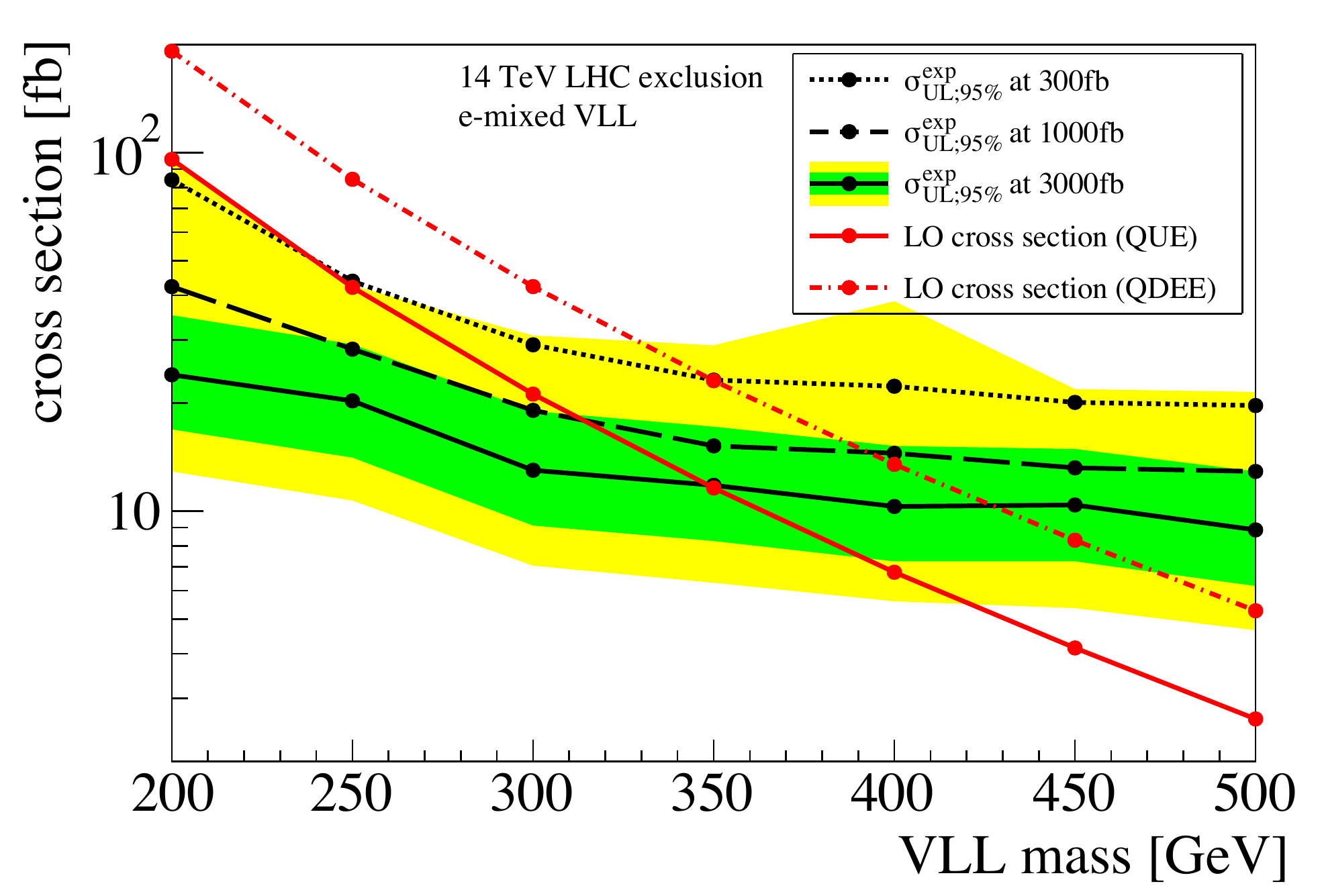} \
\includegraphics[width=.48\linewidth]{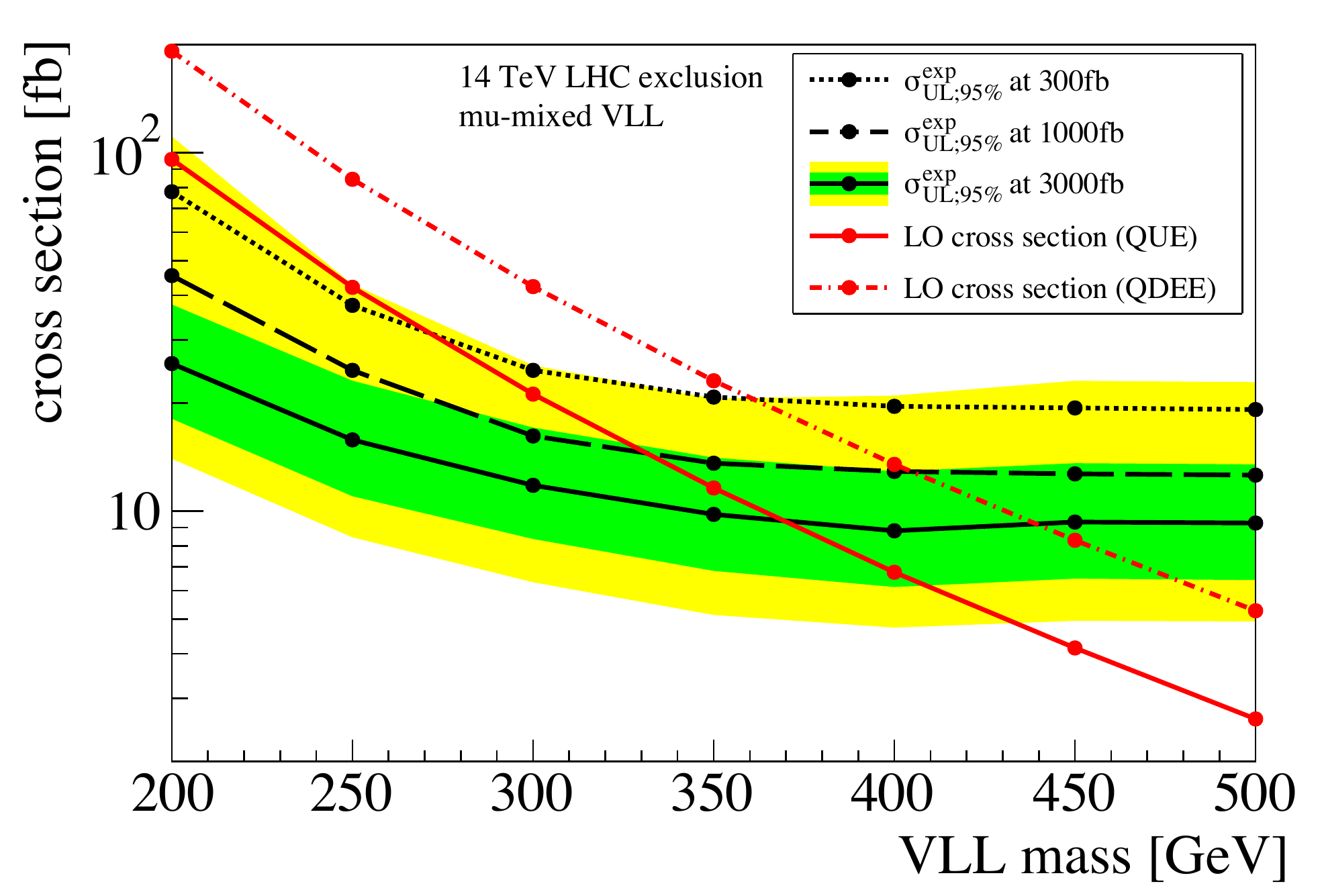}
 \caption{\label{fig:vll-exclusion}The 95\% CL expected upper limit on the production cross section of $pp\to\tau_4^+\tau_4^-$ at the LHC with $\sqrt{s}=14~\tev$ and an integrated luminosity of $\int \mathcal L =300$, 1000, and $3000\ifb$. In the left (right) plot, $\tau_4$ is assumed to be mixed exclusively with electrons (muons). The uncertainty band is shown only for $\int\mathcal L=3000\ifb$. A systematic uncertainty of 20\% is assumed for the background, and statistical uncertainty is included. The signal cross section is calculated at tree level and the theoretical uncertainty on that is not considered.}
\end{figure}

\subsection{Results}

The selection flow for the background events is summarized in \tableref{vll-BKG}.\footnote{According to the categorization of the Snowmass background set, ``di-boson'' corresponds to the sum of {\tt LLB} and {\tt BB}, ``tri-boson'' to {\tt BBB}, and ``top'' is the sum of the categories {\tt tB}, {\tt tj}, {\tt tt}, and {\tt ttB}.}
From the expected background contribution, the expected 95\% confidence level (CL) upper limit on the number of events, $N_{\rm UL}$, is calculated for each signal region with the ${\rm CL}_s$ method~\cite{Read:2002hq}, and shown in the table for three values of an integrated luminosity, $\int\mathcal L=300$, 1000, and $3000~\ifb$.
Here, we use a systematic uncertainty of 20\% for the background contributions.

Seven model points with $m_{\tau_{4}}=200\text{--}500~\gev$ are defined for both the $e$-mixed case and the $\mu$-mixed case. The selection flow of the signal events is shown in \tableref{vll-signal}. The values in this table are for the QUE model; for the QDEE model, due to the unified-mass assumptions, all the values in the table are doubled.

For each model point, the expected 95\% CL upper limit on the signal cross section, $\sigma_{\rm UL}$, is obtained by the following procedure. First, the upper limit on the signal cross section is calculated for each SR based on $N_{\rm UL}$ and the signal yield.  Then, we select the SR that gives the lowest upper limit as the most sensitive. They are indicated in \tableref{vll-signal}.  Because the SRs are not mutually exclusive, $\sigma_{\rm UL}$ for the model point is given by the most sensitive SR.

The obtained $\sigma_{\rm UL}$ are compared against the signal cross section, $\sigma(pp\to\tau^+_{4(,5)}\tau^-_{4(,5)})$, as depicted in \figref{vll-exclusion}.
The red solid (dash-dotted) lines are the signal production cross section in the QUE (QDEE) model. They are calculated at the leading order, and theoretical uncertainty is not considered for simplicity.
The black lines are the $\sigma_{\rm UL}$ at the three values of an integrated luminosity.\footnote{To be precise, the values of $\sigma_{\rm UL}$ displayed in the figures are calculated for the QUE model. The upper limits for QDEE model points are slightly better because of our statistical treatment but the difference is negligible.}
For $\int\mathcal L=3000~\ifb$, the green and yellow bands indicating the uncertainty of $\sigma_{\rm UL}$ are also displayed; the observed limits would fall in the green (yellow) band with a probability of $68\%$ ($95\%$).
Based on this comparison, the expected upper bound on the vector-like leptons are obtained for each of the four scenarios, i.e., the QUE and QDEE models with the vector-like lepton mixed with electron and muons.

The discovery sensitivity of the 14 TeV LHC is also calculated in terms of CL$_b$, i.e., $p$-value of the background-only hypothesis, as shown in \figref{vll-discovery}. Solid (dotted) lines are for $e$-mixed vector-like lepton(s) in the QUE (QDEE) model with three values of the integrated luminosity, $\int\mathcal L = 300, 1000$ and $3000~\ifb$ from top to bottom. Similar sensitivities are obtained for the $\mu$-mixed case.

The results are summarized in \tableref{vll-massreach} of the main text.

\begin{figure}[t]
\includegraphics[width=.60\linewidth]{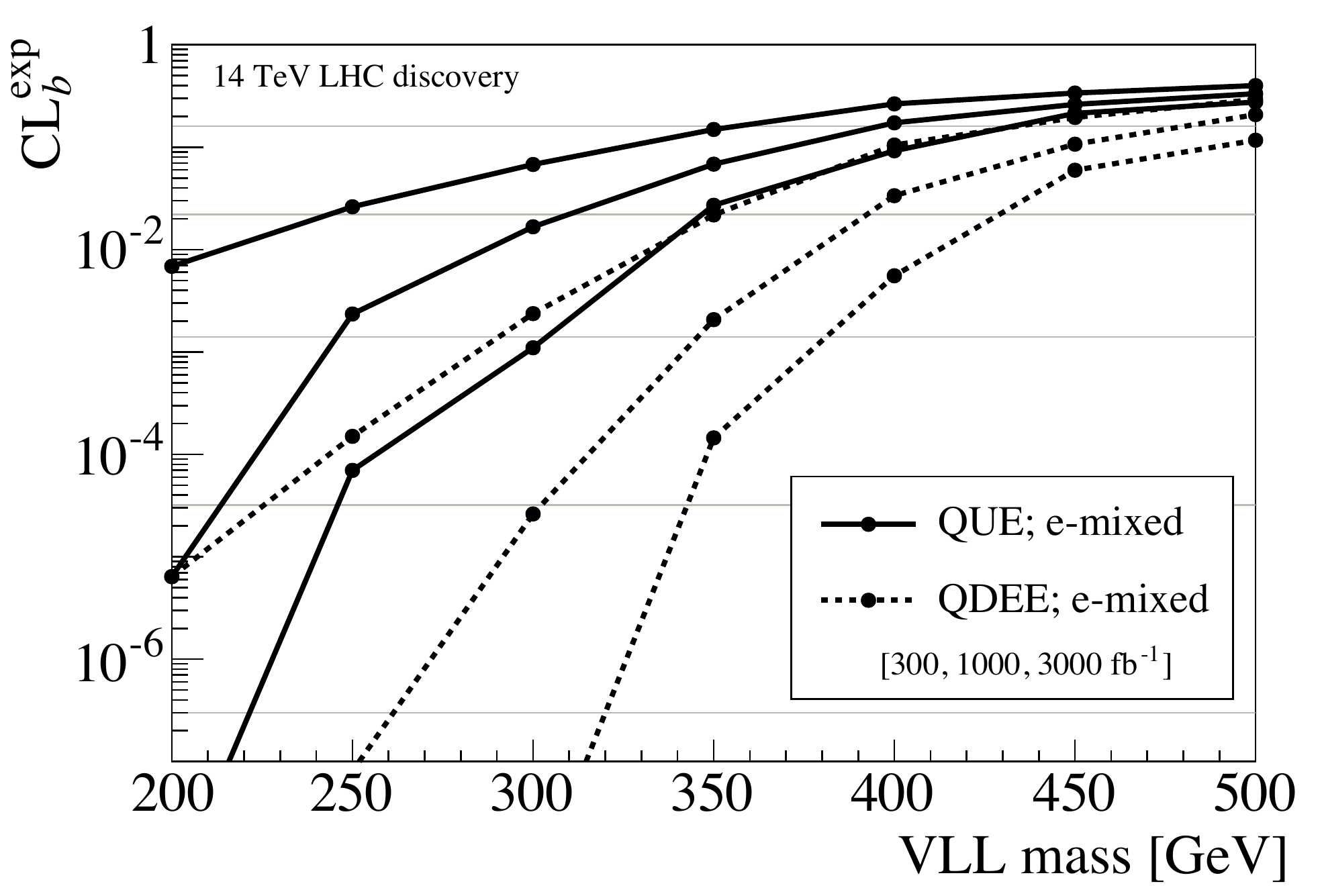}
 \caption{\label{fig:vll-discovery}The expected sensitivity of the 14 TeV LHC to the discovery of vector-like leptons, calculated under the assumption that the background contribution has a systematic uncertainty of 20\%.
Solid (dashed) lines are for QUE (QDEE) model with $e$-mixed vector-like leptons, corresponding to the integrated luminosity of $300, 1000$, and $3000~\ifb$ from top to bottom.
Similar sensitivity is expected for $\mu$-mixing cases.}
\end{figure}

\bibliography{bino4g02}

\providecommand{\href}[2]{#2}\begingroup\raggedright\begin{thebibliography}{10}

\bibitem{Feng:2013pwa}
J.~L. Feng, ``{Naturalness and the Status of Supersymmetry},''
  \href{http://dx.doi.org/10.1146/annurev-nucl-102010-130447}{{\em Ann. Rev.
  Nucl. Part. Sci.} {\bfseries 63} (2013) 351--382},
\href{http://arxiv.org/abs/1302.6587}{{\ttfamily arXiv:1302.6587 [hep-ph]}}.

\bibitem{Craig:2013cxa}
N.~Craig, ``{The State of Supersymmetry after Run I of the LHC},'' in {\em
  {Beyond the Standard Model after the first run of the LHC Arcetri, Florence,
  Italy, May 20-July 12, 2013}}.
\newblock 2013.
\newblock
\href{http://arxiv.org/abs/1309.0528}{{\ttfamily arXiv:1309.0528 [hep-ph]}}.
\newblock

\bibitem{Moroi:1991mg}
T.~Moroi and Y.~Okada, ``{Radiative corrections to Higgs masses in the
  supersymmetric model with an extra family and antifamily},''
\href{http://dx.doi.org/10.1142/S0217732392000124}{{\em Mod. Phys. Lett.}
  {\bfseries A7} (1992) 187--200}.

\bibitem{Moroi:1992zk}
T.~Moroi and Y.~Okada, ``{Upper bound of the lightest neutral Higgs mass in
  extended supersymmetric Standard Models},''
\href{http://dx.doi.org/10.1016/0370-2693(92)90091-H}{{\em Phys. Lett.}
  {\bfseries B295} (1992) 73--78}.

\bibitem{Martin:2009bg}
S.~P. Martin, ``{Extra vector-like matter and the lightest Higgs scalar boson
  mass in low-energy supersymmetry},''
  \href{http://dx.doi.org/10.1103/PhysRevD.81.035004}{{\em Phys. Rev.}
  {\bfseries D81} (2010) 035004},
\href{http://arxiv.org/abs/0910.2732}{{\ttfamily arXiv:0910.2732 [hep-ph]}}.

\bibitem{Abdullah:2015zta}
M.~Abdullah and J.~L. Feng, ``{Reviving bino dark matter with vectorlike fourth
  generation particles},''
  \href{http://dx.doi.org/10.1103/PhysRevD.93.015006}{{\em Phys. Rev.}
  {\bfseries D93} (2016) 015006},
\href{http://arxiv.org/abs/1510.06089}{{\ttfamily arXiv:1510.06089 [hep-ph]}}.

\bibitem{Kumar:2015tna}
N.~Kumar and S.~P. Martin, ``{Vectorlike leptons at the Large Hadron
  Collider},'' \href{http://dx.doi.org/10.1103/PhysRevD.92.115018}{{\em Phys.
  Rev.} {\bfseries D92} (2015) 115018},
\href{http://arxiv.org/abs/1510.03456}{{\ttfamily arXiv:1510.03456 [hep-ph]}}.

\bibitem{Drees:1993bu}
M.~Drees and M.~Nojiri, ``{Neutralino - nucleon scattering revisited},''
  \href{http://dx.doi.org/10.1103/PhysRevD.48.3483}{{\em Phys. Rev.} {\bfseries
  D48} (1993) 3483--3501},
\href{http://arxiv.org/abs/hep-ph/9307208}{{\ttfamily arXiv:hep-ph/9307208
  [hep-ph]}}.

\bibitem{Belanger:2014vza}
G.~Bélanger, F.~Boudjema, A.~Pukhov, and A.~Semenov, ``{micrOMEGAs4.1: two
  dark matter candidates},''
  \href{http://dx.doi.org/10.1016/j.cpc.2015.03.003}{{\em Comput. Phys.
  Commun.} {\bfseries 192} (2015) 322--329},
\href{http://arxiv.org/abs/1407.6129}{{\ttfamily arXiv:1407.6129 [hep-ph]}}.

\bibitem{Junnarkar:2013ac}
P.~Junnarkar and A.~Walker-Loud, ``{Scalar strange content of the nucleon from
  lattice QCD},'' \href{http://dx.doi.org/10.1103/PhysRevD.87.114510}{{\em
  Phys. Rev.} {\bfseries D87} (2013) 114510},
\href{http://arxiv.org/abs/1301.1114}{{\ttfamily arXiv:1301.1114 [hep-lat]}}.

\bibitem{Alarcon:2012nr}
J.~M. Alarcon, L.~S. Geng, J.~Martin~Camalich, and J.~A. Oller, ``{The
  strangeness content of the nucleon from effective field theory and
  phenomenology},''
  \href{http://dx.doi.org/10.1016/j.physletb.2014.01.065}{{\em Phys. Lett.}
  {\bfseries B730} (2014) 342--346},
\href{http://arxiv.org/abs/1209.2870}{{\ttfamily arXiv:1209.2870 [hep-ph]}}.

\bibitem{Alarcon:2011zs}
J.~M. Alarcon, J.~Martin~Camalich, and J.~A. Oller, ``{The chiral
  representation of the $\pi N$ scattering amplitude and the pion-nucleon sigma
  term},'' \href{http://dx.doi.org/10.1103/PhysRevD.85.051503}{{\em Phys. Rev.}
  {\bfseries D85} (2012) 051503},
\href{http://arxiv.org/abs/1110.3797}{{\ttfamily arXiv:1110.3797 [hep-ph]}}.

\bibitem{Hoferichter:2015dsa}
M.~Hoferichter, J.~Ruiz~de Elvira, B.~Kubis, and U.-G. Meißner,
  ``{High-Precision Determination of the Pion-Nucleon $\sigma$ Term from
  Roy-Steiner Equations},''
  \href{http://dx.doi.org/10.1103/PhysRevLett.115.092301}{{\em Phys. Rev.
  Lett.} {\bfseries 115} (2015) 092301},
\href{http://arxiv.org/abs/1506.04142}{{\ttfamily arXiv:1506.04142 [hep-ph]}}.

\bibitem{Belanger:2004yn}
G.~Belanger, F.~Boudjema, A.~Pukhov, and A.~Semenov, ``{micrOMEGAs: Version
  1.3},'' \href{http://dx.doi.org/10.1016/j.cpc.2005.12.005}{{\em Comput. Phys.
  Commun.} {\bfseries 174} (2006) 577--604},
\href{http://arxiv.org/abs/hep-ph/0405253}{{\ttfamily arXiv:hep-ph/0405253
  [hep-ph]}}.

\bibitem{Akerib:2015rjg}
{\bfseries LUX} Collaboration, D.~S. Akerib {\em et al.}, ``{Improved Limits on
  Scattering of Weakly Interacting Massive Particles from Reanalysis of 2013
  LUX Data},'' \href{http://dx.doi.org/10.1103/PhysRevLett.116.161301}{{\em
  Phys. Rev. Lett.} {\bfseries 116} (2016) 161301},
\href{http://arxiv.org/abs/1512.03506}{{\ttfamily arXiv:1512.03506
  [astro-ph.CO]}}.

\bibitem{Amaudruz:2014nsa}
{\bfseries DEAP} Collaboration, P.~A. Amaudruz {\em et al.},
  \href{http://dx.doi.org/10.1016/j.nuclphysbps.2015.09.048}{``{DEAP-3600 Dark
  Matter Search},''} in {\em {Proceedings, 37th International Conference on
  High Energy Physics (ICHEP 2014)}}, vol.~273-275, pp.~340--346.
\newblock 2016.
\newblock
\href{http://arxiv.org/abs/1410.7673}{{\ttfamily arXiv:1410.7673
  [physics.ins-det]}}.
\newblock

\bibitem{Aprile:2015uzo}
{\bfseries XENON} Collaboration, E.~Aprile {\em et al.}, ``{Physics reach of
  the XENON1T dark matter experiment},''
  \href{http://dx.doi.org/10.1088/1475-7516/2016/04/027}{{\em JCAP} {\bfseries
  1604} (2016) 027},
\href{http://arxiv.org/abs/1512.07501}{{\ttfamily arXiv:1512.07501
  [physics.ins-det]}}.

\bibitem{Aalseth:2015mba}
C.~E. Aalseth {\em et al.}, ``{The DarkSide Multiton Detector for the Direct
  Dark Matter Search},''
\href{http://dx.doi.org/10.1155/2015/541362}{{\em Adv. High Energy Phys.}
  {\bfseries 2015} (2015) 541362}.

\bibitem{Akerib:2015cja}
{\bfseries LZ} Collaboration, D.~S. Akerib {\em et al.}, ``{LUX-ZEPLIN (LZ)
  Conceptual Design Report},''
\href{http://arxiv.org/abs/1509.02910}{{\ttfamily arXiv:1509.02910
  [physics.ins-det]}}.

\bibitem{Aalbers:2016jon}
{\bfseries DARWIN} Collaboration, J.~Aalbers {\em et al.}, ``{DARWIN: towards
  the ultimate dark matter detector},''
\href{http://arxiv.org/abs/1606.07001}{{\ttfamily arXiv:1606.07001
  [astro-ph.IM]}}.

\bibitem{Feng:2014uja}
J.~L. Feng {\em et al.}, ``{Planning the Future of U.S. Particle Physics
  (Snowmass 2013): Chapter 4: Cosmic Frontier},'' in {\em {Community Summer
  Study 2013: Snowmass on the Mississippi (CSS2013) Minneapolis, MN, USA, July
  29-August 6, 2013}}.
\newblock 2014.
\newblock
\href{http://arxiv.org/abs/1401.6085}{{\ttfamily arXiv:1401.6085 [hep-ex]}}.
\newblock

\bibitem{Amole:2015lsj}
{\bfseries PICO} Collaboration, C.~Amole {\em et al.}, ``{Dark Matter Search
  Results from the PICO-2L C$_3$F$_8$ Bubble Chamber},''
  \href{http://dx.doi.org/10.1103/PhysRevLett.114.231302}{{\em Phys. Rev.
  Lett.} {\bfseries 114} no.~23, (2015) 231302},
\href{http://arxiv.org/abs/1503.00008}{{\ttfamily arXiv:1503.00008
  [astro-ph.CO]}}.

\bibitem{Amole:2015pla}
{\bfseries PICO} Collaboration, C.~Amole {\em et al.}, ``{Dark matter search
  results from the PICO-60 CF$_3$I bubble chamber},''
  \href{http://dx.doi.org/10.1103/PhysRevD.93.052014}{{\em Phys. Rev.}
  {\bfseries D93} no.~5, (2016) 052014},
\href{http://arxiv.org/abs/1510.07754}{{\ttfamily arXiv:1510.07754 [hep-ex]}}.

\bibitem{Aartsen:2016exj}
{\bfseries IceCube} Collaboration, M.~G. Aartsen {\em et al.}, ``{Improved
  limits on dark matter annihilation in the Sun with the 79-string IceCube
  detector and implications for supersymmetry},''
  \href{http://dx.doi.org/10.1088/1475-7516/2016/04/022}{{\em JCAP} {\bfseries
  1604} no.~04, (2016) 022},
\href{http://arxiv.org/abs/1601.00653}{{\ttfamily arXiv:1601.00653 [hep-ph]}}.

\bibitem{Aprile:2013doa}
{\bfseries XENON100} Collaboration, E.~Aprile {\em et al.}, ``{Limits on
  spin-dependent WIMP-nucleon cross sections from 225 live days of XENON100
  data},'' \href{http://dx.doi.org/10.1103/PhysRevLett.111.021301}{{\em Phys.
  Rev. Lett.} {\bfseries 111} no.~2, (2013) 021301},
\href{http://arxiv.org/abs/1301.6620}{{\ttfamily arXiv:1301.6620
  [astro-ph.CO]}}.

\bibitem{Akerib:2016lao}
{\bfseries LUX} Collaboration, D.~S. Akerib {\em et al.}, ``{Results on the
  Spin-Dependent Scattering of Weakly Interacting Massive Particles on Nucleons
  from the Run 3 Data of the LUX Experiment},''
  \href{http://dx.doi.org/10.1103/PhysRevLett.116.161302}{{\em Phys. Rev.
  Lett.} {\bfseries 116} no.~16, (2016) 161302},
\href{http://arxiv.org/abs/1602.03489}{{\ttfamily arXiv:1602.03489 [hep-ex]}}.

\bibitem{Schumann:2015cpa}
M.~Schumann, L.~Baudis, L.~Bütikofer, A.~Kish, and M.~Selvi, ``{Dark matter
  sensitivity of multi-ton liquid xenon detectors},''
  \href{http://dx.doi.org/10.1088/1475-7516/2015/10/016}{{\em JCAP} {\bfseries
  1510} (2015) 016},
\href{http://arxiv.org/abs/1506.08309}{{\ttfamily arXiv:1506.08309
  [physics.ins-det]}}.

\bibitem{Ade:2015xua}
{\bfseries Planck} Collaboration, P.~A.~R. Ade {\em et al.}, ``{Planck 2015
  results. XIII. Cosmological parameters},''
\href{http://arxiv.org/abs/1502.01589}{{\ttfamily arXiv:1502.01589
  [astro-ph.CO]}}.

\bibitem{Ahnen:2016qkx}
{\bfseries Fermi-LAT, MAGIC} Collaboration, M.~L. Ahnen {\em et al.}, ``{Limits
  to dark matter annihilation cross-section from a combined analysis of MAGIC
  and Fermi-LAT observations of dwarf satellite galaxies},''
  \href{http://dx.doi.org/10.1088/1475-7516/2016/02/039}{{\em JCAP} {\bfseries
  1602} (2016) 039},
\href{http://arxiv.org/abs/1601.06590}{{\ttfamily arXiv:1601.06590
  [astro-ph.HE]}}.

\bibitem{Carr:2015hta}
{\bfseries CTA Consortium} Collaboration, J.~Carr {\em et al.}, ``{Prospects
  for Indirect Dark Matter Searches with the Cherenkov Telescope Array
  (CTA)},'' in {\em {Proceedings, 34th International Cosmic Ray Conference
  (ICRC 2015)}}.
\newblock 2015.
\newblock
\href{http://arxiv.org/abs/1508.06128}{{\ttfamily arXiv:1508.06128
  [astro-ph.HE]}}.
\newblock

\bibitem{Abazajian:2015raa}
K.~N. Abazajian and R.~E. Keeley, ``{Bright gamma-ray Galactic Center excess
  and dark dwarfs: Strong tension for dark matter annihilation despite Milky
  Way halo profile and diffuse emission uncertainties},''
  \href{http://dx.doi.org/10.1103/PhysRevD.93.083514}{{\em Phys. Rev.}
  {\bfseries D93} no.~8, (2016) 083514},
\href{http://arxiv.org/abs/1510.06424}{{\ttfamily arXiv:1510.06424 [hep-ph]}}.

\bibitem{Ackermann:2015zua}
{\bfseries Fermi-LAT} Collaboration, M.~Ackermann {\em et al.}, ``{Searching
  for Dark Matter Annihilation from Milky Way Dwarf Spheroidal Galaxies with
  Six Years of Fermi Large Area Telescope Data},''
  \href{http://dx.doi.org/10.1103/PhysRevLett.115.231301}{{\em Phys. Rev.
  Lett.} {\bfseries 115} no.~23, (2015) 231301},
\href{http://arxiv.org/abs/1503.02641}{{\ttfamily arXiv:1503.02641
  [astro-ph.HE]}}.

\bibitem{Bonnivard:2015xpq}
V.~Bonnivard {\em et al.}, ``{Dark matter annihilation and decay in dwarf
  spheroidal galaxies: The classical and ultrafaint dSphs},''
  \href{http://dx.doi.org/10.1093/mnras/stv1601}{{\em Mon. Not. Roy. Astron.
  Soc.} {\bfseries 453} no.~1, (2015) 849--867},
\href{http://arxiv.org/abs/1504.02048}{{\ttfamily arXiv:1504.02048
  [astro-ph.HE]}}.

\bibitem{Aartsen:2015xej}
{\bfseries IceCube} Collaboration, M.~G. Aartsen {\em et al.}, ``{Search for
  Dark Matter Annihilation in the Galactic Center with IceCube-79},''
  \href{http://dx.doi.org/10.1140/epjc/s10052-015-3713-1}{{\em Eur. Phys. J.}
  {\bfseries C75} (2015) 492},
\href{http://arxiv.org/abs/1505.07259}{{\ttfamily arXiv:1505.07259
  [astro-ph.HE]}}.

\bibitem{DiMauro:2015jxa}
M.~Di~Mauro, F.~Donato, N.~Fornengo, and A.~Vittino, ``{Dark matter vs.
  astrophysics in the interpretation of AMS-02 electron and positron data},''
  \href{http://dx.doi.org/10.1088/1475-7516/2016/05/031}{{\em JCAP} {\bfseries
  1605} (2016) 031},
\href{http://arxiv.org/abs/1507.07001}{{\ttfamily arXiv:1507.07001
  [astro-ph.HE]}}.

\bibitem{Coutinho:1998bu}
Y.~D.~A. Coutinho, J.~A. Martins~Simoes, C.~M. Porto, and P.~P. Queiroz~Filho,
  ``{Single heavy lepton production in hadron hadron collisions},''
\href{http://dx.doi.org/10.1103/PhysRevD.57.6975}{{\em Phys. Rev.} {\bfseries
  D57} (1998) 6975--6980}.

\bibitem{Chatrchyan:2013oca}
{\bfseries CMS} Collaboration, S.~Chatrchyan {\em et al.}, ``{Searches for
  long-lived charged particles in pp collisions at $\sqrt{s}$=7 and 8 TeV},''
  \href{http://dx.doi.org/10.1007/JHEP07(2013)122}{{\em JHEP} {\bfseries 07}
  (2013) 122},
\href{http://arxiv.org/abs/1305.0491}{{\ttfamily arXiv:1305.0491 [hep-ex]}}.

\bibitem{CMS-PAS-EXO-15-010}
{CMS Collaboration}, ``{Searches for Long-lived Charged Particles in
  Proton-Proton Collisions at $\sqrt{s}=13$ TeV},''.
\href{https://cds.cern.ch/record/2114818}{CMS-PAS-EXO-15-010}.

\bibitem{ATLAS:2014fka}
{\bfseries ATLAS} Collaboration, G.~Aad {\em et al.}, ``{Searches for heavy
  long-lived charged particles with the ATLAS detector in proton-proton
  collisions at $ \sqrt{s}=8 $ TeV},''
  \href{http://dx.doi.org/10.1007/JHEP01(2015)068}{{\em JHEP} {\bfseries 01}
  (2015) 068},
\href{http://arxiv.org/abs/1411.6795}{{\ttfamily arXiv:1411.6795 [hep-ex]}}.

\bibitem{Feng:2015wqa}
J.~L. Feng, S.~Iwamoto, Y.~Shadmi, and S.~Tarem, ``{Long-Lived Sleptons at the
  LHC and a 100 TeV Proton Collider},''
  \href{http://dx.doi.org/10.1007/JHEP12(2015)166}{{\em JHEP} {\bfseries 12}
  (2015) 166},
\href{http://arxiv.org/abs/1505.02996}{{\ttfamily arXiv:1505.02996 [hep-ph]}}.

\bibitem{Aad:2015qfa}
{\bfseries ATLAS} Collaboration, G.~Aad {\em et al.}, ``{Search for metastable
  heavy charged particles with large ionisation energy loss in pp collisions at
  $\sqrt{s} = 8$ TeV using the ATLAS experiment},''
  \href{http://dx.doi.org/10.1140/epjc/s10052-015-3609-0}{{\em Eur. Phys. J.}
  {\bfseries C75} (2015) 407},
\href{http://arxiv.org/abs/1506.05332}{{\ttfamily arXiv:1506.05332 [hep-ex]}}.

\bibitem{Aad:2015dha}
{\bfseries ATLAS} Collaboration, G.~Aad {\em et al.}, ``{Search for heavy
  lepton resonances decaying to a $Z$ boson and a lepton in $pp$ collisions at
  $\sqrt{s}=8$ TeV with the ATLAS detector},''
  \href{http://dx.doi.org/10.1007/JHEP09(2015)108}{{\em JHEP} {\bfseries 09}
  (2015) 108},
\href{http://arxiv.org/abs/1506.01291}{{\ttfamily arXiv:1506.01291 [hep-ex]}}.

\bibitem{Falkowski:2013jya}
A.~Falkowski, D.~M. Straub, and A.~Vicente, ``{Vector-like leptons: Higgs
  decays and collider phenomenology},''
  \href{http://dx.doi.org/10.1007/JHEP05(2014)092}{{\em JHEP} {\bfseries 05}
  (2014) 092},
\href{http://arxiv.org/abs/1312.5329}{{\ttfamily arXiv:1312.5329 [hep-ph]}}.

\bibitem{Chatrchyan:2014aea}
{\bfseries CMS} Collaboration, S.~Chatrchyan {\em et al.}, ``{Search for
  anomalous production of events with three or more leptons in $pp$ collisions
  at $\sqrt(s) =$ 8 TeV},''
  \href{http://dx.doi.org/10.1103/PhysRevD.90.032006}{{\em Phys. Rev.}
  {\bfseries D90} (2014) 032006},
\href{http://arxiv.org/abs/1404.5801}{{\ttfamily arXiv:1404.5801 [hep-ex]}}.

\bibitem{Dermisek:2014qca}
R.~Dermisek, J.~P. Hall, E.~Lunghi, and S.~Shin, ``{Limits on Vectorlike
  Leptons from Searches for Anomalous Production of Multi-Lepton Events},''
  \href{http://dx.doi.org/10.1007/JHEP12(2014)013}{{\em JHEP} {\bfseries 12}
  (2014) 013},
\href{http://arxiv.org/abs/1408.3123}{{\ttfamily arXiv:1408.3123 [hep-ph]}}.

\bibitem{Achard:2001qw}
{\bfseries L3} Collaboration, P.~Achard {\em et al.}, ``{Search for heavy
  neutral and charged leptons in $e^{+} e^{-}$ annihilation at LEP},''
  \href{http://dx.doi.org/10.1016/S0370-2693(01)01005-X}{{\em Phys. Lett.}
  {\bfseries B517} (2001) 75--85},
\href{http://arxiv.org/abs/hep-ex/0107015}{{\ttfamily arXiv:hep-ex/0107015
  [hep-ex]}}.

\bibitem{DePree:2008st}
E.~De~Pree, M.~Sher, and I.~Turan, ``{Production of single heavy charged
  leptons at a linear collider},''
  \href{http://dx.doi.org/10.1103/PhysRevD.77.093001}{{\em Phys. Rev.}
  {\bfseries D77} (2008) 093001},
\href{http://arxiv.org/abs/0803.0996}{{\ttfamily arXiv:0803.0996 [hep-ph]}}.

\bibitem{Djouadi:2016eyy}
A.~Djouadi, J.~Ellis, R.~Godbole, and J.~Quevillon, ``{Future Collider
  Signatures of the Possible 750 GeV State},''
  \href{http://dx.doi.org/10.1007/JHEP03(2016)205}{{\em JHEP} {\bfseries 03}
  (2016) 205},
\href{http://arxiv.org/abs/1601.03696}{{\ttfamily arXiv:1601.03696 [hep-ph]}}.

\bibitem{Aad:2014vma}
{\bfseries ATLAS} Collaboration, G.~Aad {\em et al.}, ``{Search for direct
  production of charginos, neutralinos and sleptons in final states with two
  leptons and missing transverse momentum in $pp$ collisions at $\sqrt{s} =$ 8
  TeV with the ATLAS detector},''
  \href{http://dx.doi.org/10.1007/JHEP05(2014)071}{{\em JHEP} {\bfseries 05}
  (2014) 071},
\href{http://arxiv.org/abs/1403.5294}{{\ttfamily arXiv:1403.5294 [hep-ex]}}.

\bibitem{Khachatryan:2014qwa}
{\bfseries CMS} Collaboration, V.~Khachatryan {\em et al.}, ``{Searches for
  electroweak production of charginos, neutralinos, and sleptons decaying to
  leptons and W, Z, and Higgs bosons in pp collisions at 8 TeV},''
  \href{http://dx.doi.org/10.1140/epjc/s10052-014-3036-7}{{\em Eur. Phys. J.}
  {\bfseries C74} (2014) 3036},
\href{http://arxiv.org/abs/1405.7570}{{\ttfamily arXiv:1405.7570 [hep-ex]}}.

\bibitem{Eckel:2014dza}
J.~Eckel, M.~J. Ramsey-Musolf, W.~Shepherd, and S.~Su, ``{Impact of LSP
  Character on Slepton Reach at the LHC},''
  \href{http://dx.doi.org/10.1007/JHEP11(2014)117}{{\em JHEP} {\bfseries 11}
  (2014) 117},
\href{http://arxiv.org/abs/1408.2841}{{\ttfamily arXiv:1408.2841 [hep-ph]}}.

\bibitem{Alwall:2014hca}
J.~Alwall, R.~Frederix, S.~Frixione, V.~Hirschi, F.~Maltoni, O.~Mattelaer,
  H.~S. Shao, T.~Stelzer, P.~Torrielli, and M.~Zaro, ``{The automated
  computation of tree-level and next-to-leading order differential cross
  sections, and their matching to parton shower simulations},''
  \href{http://dx.doi.org/10.1007/JHEP07(2014)079}{{\em JHEP} {\bfseries 07}
  (2014) 079},
\href{http://arxiv.org/abs/1405.0301}{{\ttfamily arXiv:1405.0301 [hep-ph]}}.

\bibitem{Pythia6.4}
T.~Sjostrand, S.~Mrenna, and P.~Z. Skands, ``{PYTHIA 6.4 Physics and Manual},''
  \href{http://dx.doi.org/10.1088/1126-6708/2006/05/026}{{\em JHEP} {\bfseries
  05} (2006) 026},
\href{http://arxiv.org/abs/hep-ph/0603175}{{\ttfamily arXiv:hep-ph/0603175}}.

\bibitem{deFavereau:2013fsa}
{\bfseries DELPHES 3} Collaboration, J.~de~Favereau, C.~Delaere, P.~Demin,
  A.~Giammanco, V.~Lemaître, A.~Mertens, and M.~Selvaggi, ``{DELPHES 3, A
  modular framework for fast simulation of a generic collider experiment},''
  \href{http://dx.doi.org/10.1007/JHEP02(2014)057}{{\em JHEP} {\bfseries 02}
  (2014) 057},
\href{http://arxiv.org/abs/1307.6346}{{\ttfamily arXiv:1307.6346 [hep-ex]}}.

\bibitem{Cacciari:2005hq}
M.~Cacciari and G.~P. Salam, ``{Dispelling the $N^{3}$ myth for the $k_t$
  jet-finder},'' \href{http://dx.doi.org/10.1016/j.physletb.2006.08.037}{{\em
  Phys. Lett.} {\bfseries B641} (2006) 57--61},
\href{http://arxiv.org/abs/hep-ph/0512210}{{\ttfamily arXiv:hep-ph/0512210
  [hep-ph]}}.

\bibitem{Cacciari:2011ma}
M.~Cacciari, G.~P. Salam, and G.~Soyez, ``{FastJet User Manual},''
  \href{http://dx.doi.org/10.1140/epjc/s10052-012-1896-2}{{\em Eur. Phys. J.}
  {\bfseries C72} (2012) 1896},
\href{http://arxiv.org/abs/1111.6097}{{\ttfamily arXiv:1111.6097 [hep-ph]}}.

\bibitem{Aad:2014yka}
{\bfseries ATLAS} Collaboration, G.~Aad {\em et al.}, ``{Search for the direct
  production of charginos, neutralinos and staus in final states with at least
  two hadronically decaying taus and missing transverse momentum in $pp$
  collisions at $\sqrt{s}$ = 8 TeV with the ATLAS detector},''
  \href{http://dx.doi.org/10.1007/JHEP10(2014)096}{{\em JHEP} {\bfseries 10}
  (2014) 096},
\href{http://arxiv.org/abs/1407.0350}{{\ttfamily arXiv:1407.0350 [hep-ex]}}.

\bibitem{CMS-PAS-SUS-14-022}
{CMS Collaboration}, ``{Search for electroweak production of charginos in final
  states with two tau leptons in pp collisions at $\sqrt{s}=
  8~\mathrm{TeV}$},''.
\href{https://cds.cern.ch/record/2137232}{CMS-PAS-SUS-14-022}.

\bibitem{Endo:2011xq}
M.~Endo, K.~Hamaguchi, S.~Iwamoto, and N.~Yokozaki, ``{Higgs mass, muon g-2,
  and LHC prospects in gauge mediation models with vector-like matters},''
  \href{http://dx.doi.org/10.1103/PhysRevD.85.095012}{{\em Phys. Rev.}
  {\bfseries D85} (2012) 095012},
\href{http://arxiv.org/abs/1112.5653}{{\ttfamily arXiv:1112.5653 [hep-ph]}}.

\bibitem{Harigaya:2012ir}
K.~Harigaya, S.~Matsumoto, M.~M. Nojiri, and K.~Tobioka, ``{Search for the Top
  Partner at the LHC using Multi-b-Jet Channels},''
  \href{http://dx.doi.org/10.1103/PhysRevD.86.015005}{{\em Phys. Rev.}
  {\bfseries D86} (2012) 015005},
\href{http://arxiv.org/abs/1204.2317}{{\ttfamily arXiv:1204.2317 [hep-ph]}}.

\bibitem{Endo:2014bsa}
M.~Endo, K.~Hamaguchi, K.~Ishikawa, and M.~Stoll, ``{Reconstruction of
  Vector-like Top Partner from Fully Hadronic Final States},''
  \href{http://dx.doi.org/10.1103/PhysRevD.90.055027}{{\em Phys. Rev.}
  {\bfseries D90} no.~5, (2014) 055027},
\href{http://arxiv.org/abs/1405.2677}{{\ttfamily arXiv:1405.2677 [hep-ph]}}.

\bibitem{Mayet:2016zxu}
F.~Mayet {\em et al.}, ``{A review of the discovery reach of directional Dark
  Matter detection},''
  \href{http://dx.doi.org/10.1016/j.physrep.2016.02.007}{{\em Phys. Rept.}
  {\bfseries 627} (2016) 1--49},
\href{http://arxiv.org/abs/1602.03781}{{\ttfamily arXiv:1602.03781
  [astro-ph.CO]}}.

\bibitem{Ahlen:1987mn}
S.~P. Ahlen, F.~T. Avignone, R.~L. Brodzinski, A.~K. Drukier, G.~Gelmini, and
  D.~N. Spergel, ``{Limits on Cold Dark Matter Candidates from an Ultralow
  Background Germanium Spectrometer},''
\href{http://dx.doi.org/10.1016/0370-2693(87)91581-4}{{\em Phys. Lett.}
  {\bfseries B195} (1987) 603--608}.

\bibitem{Jungman:1995df}
G.~Jungman, M.~Kamionkowski, and K.~Griest, ``{Supersymmetric dark matter},''
  \href{http://dx.doi.org/10.1016/0370-1573(95)00058-5}{{\em Phys. Rept.}
  {\bfseries 267} (1996) 195--373},
\href{http://arxiv.org/abs/hep-ph/9506380}{{\ttfamily arXiv:hep-ph/9506380
  [hep-ph]}}.

\bibitem{Anderson:2013kxz}
J.~Anderson {\em et al.}, ``{Snowmass Energy Frontier Simulations},'' in {\em
  {Community Summer Study 2013: Snowmass on the Mississippi (CSS2013)
  Minneapolis, MN, USA, July 29-August 6, 2013}}.
\newblock 2013.
\newblock
\href{http://arxiv.org/abs/1309.1057}{{\ttfamily arXiv:1309.1057 [hep-ex]}}.
\newblock

\bibitem{Avetisyan:2013onh}
A.~Avetisyan {\em et al.}, ``{Methods and Results for Standard Model Event
  Generation at $\sqrt{s}$ = 14 TeV, 33 TeV and 100 TeV Proton Colliders (A
  Snowmass Whitepaper)},'' in {\em {Community Summer Study 2013: Snowmass on
  the Mississippi (CSS2013) Minneapolis, MN, USA, July 29-August 6, 2013}}.
\newblock 2013.
\newblock
\href{http://arxiv.org/abs/1308.1636}{{\ttfamily arXiv:1308.1636 [hep-ex]}}.
\newblock

\bibitem{Avetisyan:2013dta}
A.~Avetisyan {\em et al.}, ``{Snowmass Energy Frontier Simulations using the
  Open Science Grid (A Snowmass 2013 whitepaper)},'' in {\em {Community Summer
  Study 2013: Snowmass on the Mississippi (CSS2013) Minneapolis, MN, USA, July
  29-August 6, 2013}}.
\newblock 2013.
\newblock
\href{http://arxiv.org/abs/1308.0843}{{\ttfamily arXiv:1308.0843 [hep-ex]}}.
\newblock

\bibitem{Cacciari:2008gp}
M.~Cacciari, G.~P. Salam, and G.~Soyez, ``{The Anti-k(t) jet clustering
  algorithm},'' \href{http://dx.doi.org/10.1088/1126-6708/2008/04/063}{{\em
  JHEP} {\bfseries 04} (2008) 063},
\href{http://arxiv.org/abs/0802.1189}{{\ttfamily arXiv:0802.1189 [hep-ph]}}.

\bibitem{Aad:2014nua}
{\bfseries ATLAS} Collaboration, G.~Aad {\em et al.}, ``{Search for direct
  production of charginos and neutralinos in events with three leptons and
  missing transverse momentum in $\sqrt{s} =$ 8TeV $pp$ collisions with the
  ATLAS detector},'' \href{http://dx.doi.org/10.1007/JHEP04(2014)169}{{\em
  JHEP} {\bfseries 04} (2014) 169},
\href{http://arxiv.org/abs/1402.7029}{{\ttfamily arXiv:1402.7029 [hep-ex]}}.

\bibitem{Read:2002hq}
A.~L. Read, ``{Presentation of search results: The CL(s) technique},''
\href{http://dx.doi.org/10.1088/0954-3899/28/10/313}{{\em J. Phys.} {\bfseries
  G28} (2002) 2693--2704}.

\end{thebibliography}\endgroup

\end{document}
